%% file: main.tex
\pdfoutput=1

\documentclass[letterpaper,twocolumn,10pt]{article}
\usepackage{usenix2019_v3}
\input{customs}

\begin{document}

\date{}

\pagenumbering{gobble}

\title{\Large \bf Bots can Snoop: Uncovering and Mitigating Privacy Risks of Bots in Group Chats}

\author{
{\rm Kai-Hsiang Chou\footnotemark[1]}\\
National Taiwan University
\and
{\rm Yi-Min Lin\footnotemark[1]}\\
National Taiwan University
\and
{\rm Yi-An Wang}\\
National Taiwan University
\and
{\rm Jonathan Weiping Li }\\
National Taiwan University
\and
{\rm Tiffany Hyun-Jin Kim}\\
HRL Laboratories
\and
{\rm Hsu-Chun Hsiao\footnotemark[2]}\\
National Taiwan University
\\ Academia Sinica
} %

\maketitle

\begingroup
\renewcommand\thefootnote{\fnsymbol{footnote}}
\footnotetext[1]{Both authors contributed equally to this research.}
\footnotetext[2]{Hsu-Chun Hsiao (hchsiao@csie.ntu.edu.tw) is the corresponding author.}
\endgroup

\begin{abstract}
\input{sections/abstract}

\end{abstract}

\input{sections/introduction}

\input{sections/background}

\input{sections/attack}

\input{sections/case_studies}

\input{sections/protocol}

\input{sections/evaluation}

\input{sections/discussions}

\input{sections/related_work}

\input{sections/conclusion}

\section*{Acknowledgments}
We thank the anonymous reviewers and shepherd for their valuable comments. We also thank Mahmood Sharif and Hsien-En Tzeng for their insightful feedback.  
This research was supported in part by the National Science and Technology Council of Taiwan under grants 112-2223-E-002-010-MY4 and 113-2634-F-002-001-MBK, and National Taiwan University under grant 114L7848.

\input{sections/ethics}

\bibliographystyle{plain}
\bibliography{reference}

\input{sections/appendix}

\end{document}

%% file: customs.tex
\usepackage{multicol}
\usepackage{multirow,makecell}
\usepackage{enumitem}
\usepackage{hyperref}
\usepackage{amssymb, amsmath, amsthm}
\usepackage{pifont}
\usepackage{algorithm2e}
\usepackage{algpseudocode}
\usepackage{msc}
\usepackage{rotating}
\usepackage{lipsum}
\usepackage[operators,sets,advantage,adversary,primitives,complexity,oracles,landau,probability,logic,lambda, keys,notions,asymptotics]{cryptocode}
\usepackage{subcaption}
\usepackage{pgfplots}
\usepackage{scalefnt}
\usepackage{xspace}
\usepackage{diagbox}
\usepackage{pict2e}
\usepackage{adjustbox}
\usepackage{array}
\usepackage{xcolor}
\usepackage{tikz}
\usepackage{wasysym}
\usepackage{kantlipsum}
\usepackage[available]{usenixbadges}

\newcommand\cgka{\ensuremath{\mathsf{CGKA}}\xspace}

\newcommand\cmrtt{\ensuremath{\mathsf{CMRT}}\xspace}
\newcommand\cmrt{SnoopGuard\xspace}

\newcommand\cid{\mathsf{CID}}

\newcommand\id{\mathsf{ID}}

\newcommand{\pke}{\mathsf{PKE}}
\newcommand{\pkeg}{\mathsf{PKEG}}
\newcommand{\pkenc}{\mathsf{PKEnc}}
\newcommand{\pkdec}{\mathsf{PKDec}}

\newcommand{\add}{\mathsf{add}}
\newcommand{\creategroup}{\mathsf{create}}
\newcommand{\init}{\mathsf{init}}
\newcommand{\rem}{\mathsf{rem}}
\newcommand{\upd}{\mathsf{upd}}
\newcommand{\proc}{\mathsf{proc}}

\newcommand{\csk}{\mathsf{csk}}
\newcommand{\cpk}{\mathsf{cpk}}
\newcommand{\gpk}{\mathsf{gpk}}
\newcommand{\gsk}{\mathsf{gsk}}

\newcommand{\addcbt}{\mathsf{add}\text{--}\mathsf{cbt}}
\newcommand{\remcbt}{\mathsf{rem}\text{--}\mathsf{cbt}}

\newcommand{\getpk}{\mathsf{get}\text{--}\mathsf{pk}}
\newcommand{\setpk}{\mathsf{set}\text{--}\mathsf{pk}}

\newcommand{\usrsend}{\mathsf{usr}\text{--}\mathsf{send}}
\newcommand{\usrrecv}{\mathsf{usr}\text{--}\mathsf{recv}}
\newcommand{\cbtsend}{\mathsf{cbt}\text{--}\mathsf{send}}
\newcommand{\cbtrecv}{\mathsf{cbt}\text{--}\mathsf{recv}}

\newcommand{\me}{\mathsf{ME}}

\newcommand{\chatbots}{\mathsf{cbts}}

\newtheorem{definition}{Definition}

\definecolor{color1}{RGB}{126,122,120}
\definecolor{color2}{RGB}{7,110,202}
\definecolor{color3}{RGB}{230,25,75}

\newcommand\myparagraph[1]{\vspace{1mm}\noindent\textbf{#1}}

\newcommand*\circled[1]{\raisebox{.5pt}{\textcircled{\raisebox{-.9pt} {\small{#1}}}}}

%% file: sections/abstract.tex
New privacy concerns arise with chatbots on group messaging platforms. Chatbots may access information beyond their intended functionalities, such as sender identities or messages unintended for chatbots. Chatbot developers may exploit such information to infer personal information and link users across groups, potentially leading to data breaches, pervasive tracking, or targeted advertising.
Our analysis of conversation datasets shows that (1) chatbots often access far more messages than needed, and (2) when a user joins a new group with chatbots, there is a 3.6\% chance that at least one of the chatbots can recognize and associate the user with their previous interactions in other groups.
Although state-of-the-art (SoA) group messaging protocols provide robust end-to-end encryption and some platforms have implemented policies to limit chatbot access, no platforms successfully combine these features. 
This paper introduces \cmrt, a secure group messaging protocol that ensures user privacy against chatbots while maintaining strong end-to-end security. Our protocol offers (1) selective message access, preventing chatbots from accessing unrelated messages, and (2) sender anonymity, hiding user identities from chatbots.
\cmrt achieves $O(\log n + m)$ message-sending complexity for a group of $n$ users and $m$ chatbots, compared to $O(\log(n + m))$ in SoA protocols, with acceptable overhead for enhanced privacy. 
Our prototype implementation shows that sending a message to a group of 50 users and 10 chatbots takes about 10 milliseconds when integrated with Message Layer Security (MLS).

%% file: sections/introduction.tex
\section{Introduction}
\label{sec:Intro}

Chatbots have recently soared in popularity, partially thanks to the development of powerful Large Language Models (LLMs)~\cite{LLM_chatbot,chatbot_popularity}.
Even before the advent of LLMs, sophisticated rule-based conversational agents have been part of group messaging platforms. They provide a range of functions, from facilitating multilingual communication~\cite{yandexbot} to regulating chat content and group members~\cite{groupbutler}. 
However, this integration into group chats presents complex security challenges that go beyond the scope of typical one-on-one interactions.
Unlike one-on-one interactions, where chatbots are the direct recipients, group conversations can inadvertently expose more information than necessary for the chatbot’s functionality. For instance, granting full access to all group messages to a chatbot, which is only triggered by predefined commands, clearly violates the principle of least privilege, raising significant privacy concerns. In this work, we focus on two types of excessive information exposure:

\begin{figure}[tb]
    \centering
    \includegraphics[width=\columnwidth]{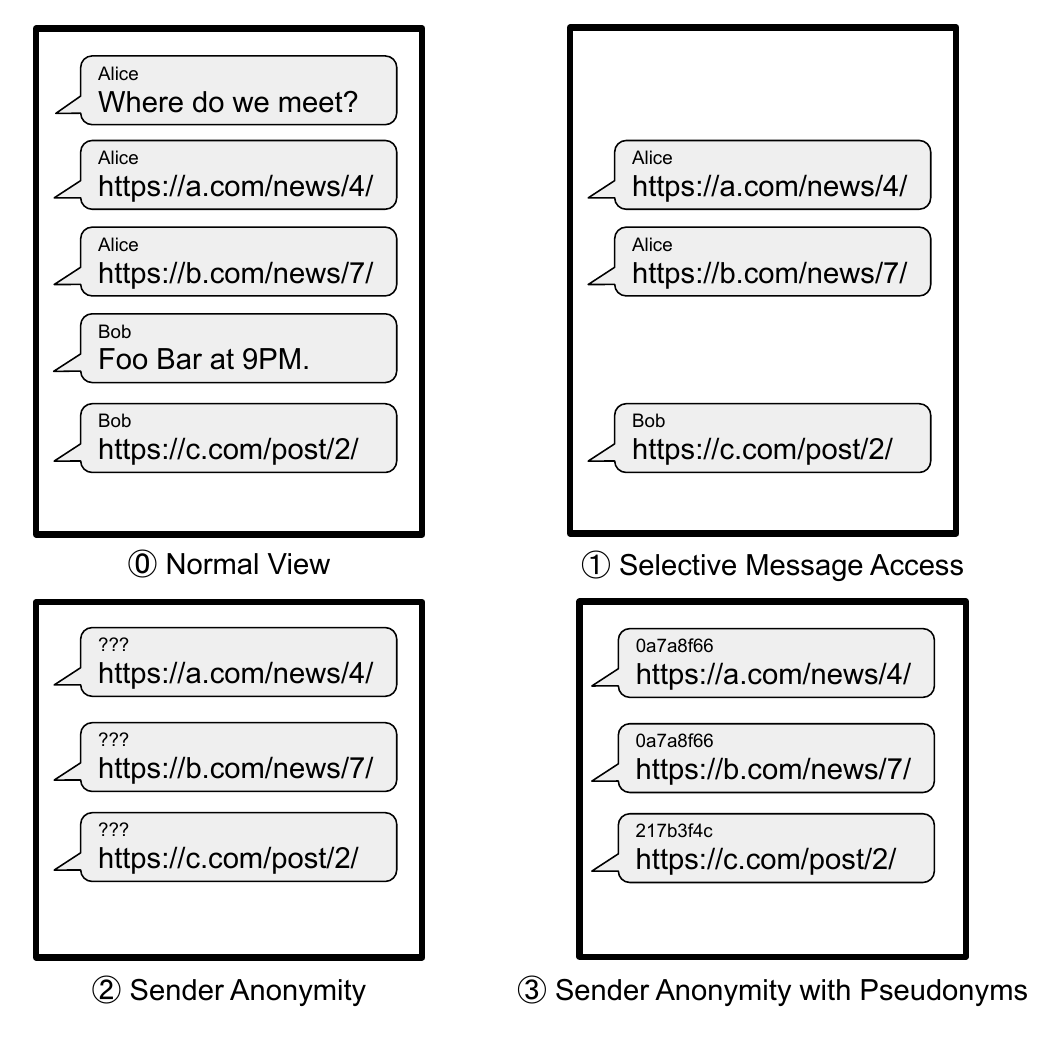}
    \caption{Example views of typical group messages from a \textit{URL-checking bot} (\circled{0}) and messages when our desired privacy policy is enforced (\circled{1}--\circled{3}). Selective message access \circled{1} ensures that the chatbot can only view messages \textit{relevant} to its functionality (i.e., containing URLs). With sender anonymity \circled{2}, the chatbot does not know the sender's identity. 
    With sender anonymity featuring pseudonyms (\circled{3}), the chatbot can distinguish between senders using pseudonymous identifiers that may change over time.}
    \label{fig:overview}
\end{figure}

\myparagraph{(1) Irrelevant Messages:} Those messages irrelevant to chatbots might contain sensitive data, including personally identifiable information (PII) like phone numbers and credit card numbers. Even in the absence of explicitly sensitive data, significant personal details can be inferred from dialogues. For instance, arranging a dinner can inadvertently reveal participants' locations. Staab et al. have shown that large language models (LLMs) can accurately infer personal attributes such as gender and location from conversations, even without explicitly mentioning these attributes~\cite{staab2023beyond}. 
These details can be used to build comprehensive profiles of group members, raising significant privacy concerns and exposing users to potential eavesdropping by seemingly harmless chatbots. Edu et al. reported a case where a chatbot developer was caught opening a URL shared in a honeypot group chat~\cite{edu2022exploring}, reinforcing the claim that chatbots can be used to gather information beyond their intended functionality.

\myparagraph{(2) User Identities.} Besides processing irrelevant messages, a chatbot developer can link private information to specific users using metadata associated with each message. For example, the Telegram chatbot ``Dr.Web'' provides a service to check for malicious links in group messages~\cite{drweb}. Since Telegram's API includes a user ID with each message, Dr.Web could potentially record users' browsing history when analyzing shared links, even across multiple groups. This risk is akin to the privacy concerns posed by ubiquitous third-party cookies, which enable trackers to covertly collect parts of users' browsing histories. In this context, the access to user metadata, which is not essential for Dr.Web's functionality, exemplifies excessive exposure of user identities to chatbots. 
This risk is exacerbated when user messages not intended for chatbots are linked to users' identities, facilitating online profiling.

To address such privacy concerns, Platforms like Telegram, Slack, and Keybase have introduced various measures to reduce the exposure of user information to chatbots. However, there is a tension between these measures and the goal of implementing end-to-end encryption (E2EE), a common standard in messaging applications that protects user privacy from service providers. Our case studies (\S\ref{sec:Case Studies}) reveal that none of the existing platforms sufficiently resolve this tension, highlighting the need for new solutions.

\subsection{Our Contributions}

This paper identifies two significant privacy issues in group messaging platforms with chatbots: (1) chatbots accessing all messages, regardless of relevance to their functions, and (2) unnecessary disclosure of message senders’ identities to chatbots.

To understand the significance of privacy concerns, we conducted case studies on two conversation datasets and show that these problems are prevalence.
First, our case study of a Discord chatbot shows that while only 0.24\% of messages are intended to the chatbot, it has access to all messages, resulting in excessive information access. Second, our analysis on a Telegram dataset shows that 3.6\% of users encounter the same chatbots in multiple groups. This highlights the prevalence of users being identified by the same chatbots across multiple groups. 
We also conducted a survey on messaging platforms to explore how they address privacy issues. By building chatbots on various platforms, we examined the consistency between their policies and implementations. None of the platforms studied mitigates privacy concerns while maintaining modern E2EE. We highlighted the tension, which our proposed protocol aims to address.

We highlight two key properties for secure group messaging: (1) Selective Message Access, which prevents chatbots from accessing irrelevant messages, and (2) Sender Anonymity, which hides user identities from chatbots, each addressing a distinct privacy issue. Figure~\ref{fig:overview} illustrates the impact of the two properties from the perspective of a chatbot, showcasing how they enhance privacy within group chats. 

To realize these critical privacy properties, we propose \cmrt, a secure group messaging protocol that supports both while ensuring strong E2EE. Our design leverages continuous group key agreement (CGKA) schemes to maintain robust security, incorporates individual key management for each chatbot to enable selective message access, and extends existing CGKA protocols to support sender anonymity. 

Theoretical analysis shows that \cmrt requires $\bigO{\log n + m}$ cryptographic operations to send a message in a group of $n$ users and $m$ chatbots, compared to the $\bigO{\log(n + m)}$ required by Message Layer Security (MLS). This demonstrates that our additional privacy protections introduce a manageable overhead for groups with a small number of chatbots, suitable for most practical scenarios. 
Our prototype implementation, which is based on Signal's Protocol and MLS, demonstrates that sending a message within a group of 50 users and 10 chatbots takes roughly 10 milliseconds on a Mac mini with an M2 processor.

%% file: sections/background.tex
\section{Background on Group Chat and Chatbot}

\subsection{Messaging Platforms}
Messaging platforms like WhatsApp, Telegram, LINE, and Slack facilitate real-time communication between users, either one-on-one or in groups. Specifically, their group (multi-user) chat functionality allows a group of people to converse asynchronously. The group initiator has the authority to select members for the group, while other users can join later. Members can send and receive messages at their convenience, without the need to be online simultaneously, as the messaging platform relays messages sent by users to other users when they are online. 

\begin{figure}[t]
    \centering
    \includegraphics[width=\linewidth]{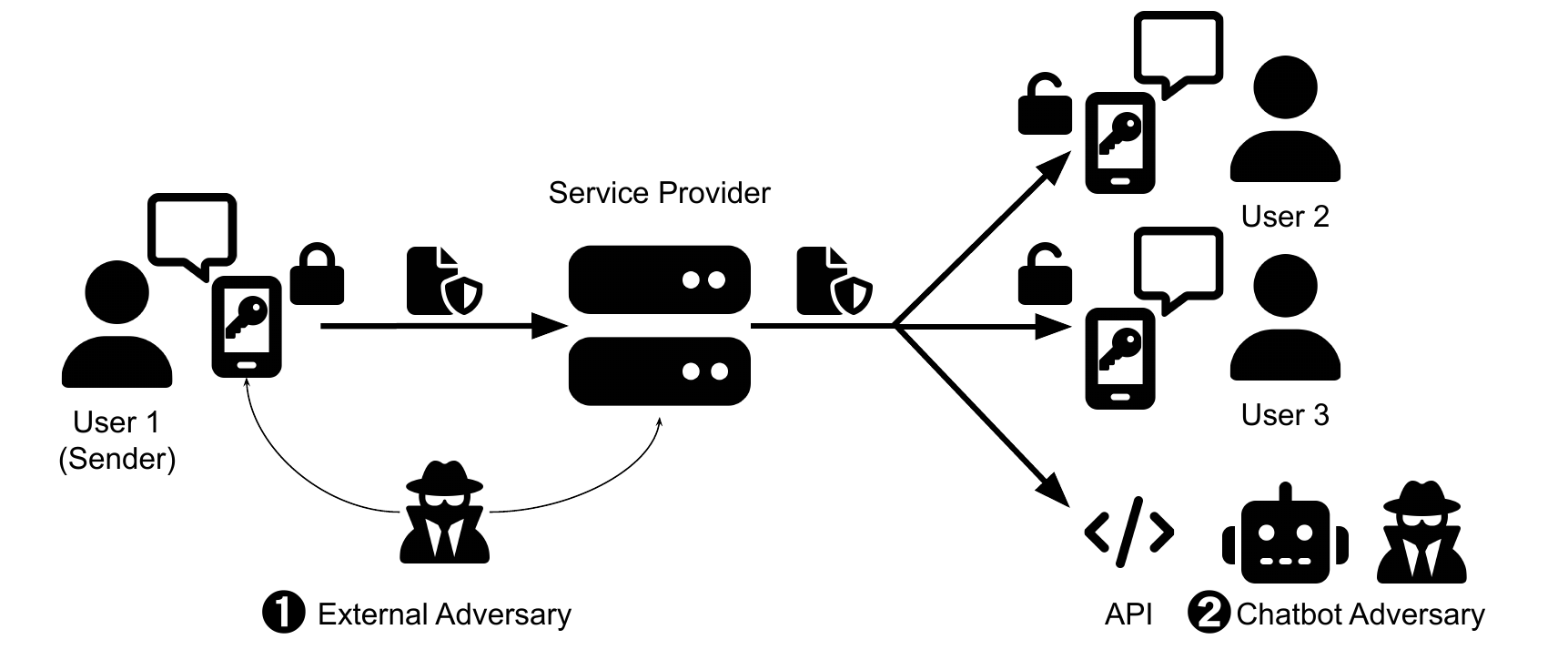}
    \caption{A typical messaging platform involves a service provider forwarding messages among group members. Two primary adversaries against users' privacy in this setting are \ding{202} malicious service providers with key-compromise capability and  \ding{203} overprivileged chatbots. While state-of-the-art secure group messaging can address \ding{202} only, our work aims to address both. The icons in the figure are from \href{https://fontawesome.com/}{Font Awesome}.}
    \label{fig:role}
\end{figure}

Many messaging platforms support adding chatbots to groups, such as Telegram, Discord, and LINE. Group members can interact with chatbots using text-based commands or natural-language prompts, enhancing group communication and productivity by providing convenient access to online services. For example, chatbots can assist multilingual groups with real-time translation~\cite{yandexbot} and help detect phishing URLs~\cite{drweb} or false information~\cite{antie_meiyu} shared by group members. Figure~\ref{fig:role} illustrates how messages are delivered within a group of four members, including users 1-3 and a chatbot. The messaging platform runs a centralized server called \textit{service provider} that facilitates the delivery of user 1's messages to both users 2 and 3, as well as to the chatbot.

\subsection{Secure Messaging}
Modern messaging applications rely on service providers, the servers of messaging platforms, to buffer and deliver messages (as shown in Figure~\ref{fig:role}).
This centralized design presents a risk, as a curious service provider could potentially eavesdrop on user messages.
Therefore, the primary goal of secure messaging protocols is to ensure that only senders and receivers can decrypt messages using end-to-end encryption (E2EE). 
E2EE guarantees that only the parties involved in the communication group can access the plaintext messages, and is now widely accepted as the security standard for messaging platforms. In addition, it is desirable that this security guarantee be resilient to key compromise.

Secure messaging protocols were initially developed for two-party communications and later extended to group settings. In two-party secure messaging, Borisov et al.~\cite{borisov2004off} highlighted the vulnerability of key compromise, showing that traditional public key cryptography, like Pretty Good Privacy (PGP), does not protect messages encrypted \textit{before} a compromise. They introduced the off-the-record (OTR) messaging, where parties continuously negotiate new Diffie-Hellman (DH) session keys and delete old keys to achieve forward secrecy (FS)~\cite{gunther1990identity}. Cohn et al.~\cite{cohn2016post} expanded the notion of FS to protect message confidentiality and integrity \textit{after} key compromise, introducing post-compromise secrecy (PCS). They demonstrated that only stateful protocols can provide PCS against full key compromise. Building on OTR, the double ratchet algorithm~\cite{perrin2016double} uses OTR’s ratcheting design to generate fresh session keys for each message. This algorithm is the foundation for key agreement in popular messaging apps like WhatsApp, Signal, and Messenger's secret conversations, and has been formally proven to ensure both FS and PCS~\cite{cohn2020formal,alwen2019double}.

To extend secure messaging from two to multiple parties, a straightforward design is to use pairwise secure channels between each two members, but the updating complexity is linear to the group size, lacking scalability for large groups. 
Another approach is the Sender Keys Protocol~\cite{WhatsApp}, which  Signal and WhatsApp use for large groups; each member generates their own encryption key called a \textit{sender key} and distributes the key to each group member through pairwise secure channels. The Sender Keys Protocol provides constant-time update and FS, but does not provide PCS~\cite{balbas2022analysis, bienstock2022more}.

Recently, continuous group key agreement (CGKA)~\cite{alwen2020security} was introduced as a unifying framework for group key agreement schemes, supporting asynchronous operations and strong security guarantees. CGKA schemes frequently update group keys to maintain confidentiality after potential key compromises, achieving FS and PCS. Tree-based CGKA protocols use key trees to reduce update complexity to logarithmic time. For example, Asynchronous Ratcheting Tree (ART)~\cite{cohn2018ends} is based on a Diffie-Hellman tree~\cite{kim2004tree}, while TreeKEM~\cite{bhargavan2018treekem} uses a hash tree for greater efficiency. Message Layer Security (MLS), an IETF standard~\cite{rfc9420}, adopts TreeKEM for group key management, and the security of MLS, TreeKEM, and related variants has been thoroughly analyzed~\cite{alwen2020security, alwen2020continuous, alwen2021modular}.
Although state-of-the-art secure group messaging, such as MLS, can defend against malicious service providers with key compromise capability (\ding{202} in Figure~\ref{fig:role}), to our knowledge, no existing protocols can protect user privacy against overprivileged chatbots as well (\ding{203} in Figure~\ref{fig:role}).

\subsection{Threat Model and Assumptions}
We consider two types of adversaries: \textit{overprivileged chatbots} and \textit{external adversaries}, as shown in Figure~\ref{fig:role}. %
External adversaries include common adversaries considered in previous literature on secure messaging~\cite{rosler2018more, unger2015sok}.

\myparagraph{Overprivileged Chatbots.} 
Overprivileged chatbots are insider adversaries participating in group conversations as regular members. They have access to the group messages, group metadata, and group events.
We assume that chatbots are passive adversaries. In other words, active attacks, such as sending malicious group modification messages~\cite{rosler2018more}, are out of scope.
While collusion between chatbots is possible, we assume no collusion between a chatbot and a group member, since such collusion would allow the chatbot to trivially access all information known to the member.

\myparagraph{External Adversaries.} 
External adversaries, including semi-honest service providers, have key compromise capabilities, allowing them to learn the keys on a user's device~\cite{cohn2016post}.
Following the typical assumption in PCS literature, we assume that a device will be compromised for only a finite period, after which the user will regain control and perform at least one secure operation.
We assume that the service provider and chatbots do not collude with each other and acknowledge that such collusion could introduce new privacy risks outside the scope of this work.

We assume that the service provider removes users' network-level metadata (such as source IP addresses) before forwarding messages to chatbots. This assumption is practical in real-world scenarios and allows us to focus solely on ensuring application-level anonymity.

%% file: sections/attack.tex
\section{Privacy Issues of Chatbots in Group Chats}
\label{sec:attack}

We present two case studies examining privacy risks from overprivileged chatbots in group settings. These studies analyze to what extent chatbots can access excessive message content and metadata beyond their functional requirements. Our analysis addresses the following research questions (RQs):

\begin{itemize}
    \item \textbf{RQ1-a:} Are chatbots overprivileged in a single group?
    \item \textbf{RQ1-b:} If so, how many unnecessary messages and sender identities can a chatbot access in a group?
    \item \textbf{RQ2:} How likely are users to be identified by the same chatbots across multiple groups?
\end{itemize}

Our case studies used two datasets: DISCO~\cite{subash2022disco} and Pushshift~\cite{baumgartner2020pushshift}. DISCO contains 1.5M Discord messages from 323.6K users across four groups (November 2019-October 2020). With readable English content but little group overlap, DISCO is better suited for analyzing excessive access in a single group (RQ1). Pushshift comprises 2.2M Telegram users across 27.8K groups (September 2015-November 2019), making it ideal for studying privacy leaks related to cross-group user linkage (RQ2).\footnote{The Pushshift dataset includes only Telegram ``channels,'' which are public chatrooms open to any user. In Telegram, ``groups'' are chatrooms restricted to approved users. For consistency, we will use the term ``group'' to refer to ``Telegram channels'' throughout this discussion.}

For ethical considerations and practical constraints, we focus on analyzing public datasets. Although these public datasets have limitations in representing private group chat dynamics, they enable a glance at potential privacy concerns, particularly regarding excessive access permissions and bot presence across multiple groups. We leave it to future work to investigate the exposure of sensitive content in private group conversations.

\subsection{Case Study 1: Unnecessary Access}
\label{sec:attack:same-group}

\myparagraph{Methodology.}
To answer RQ1, we conducted a case study of one chatbot in the DISCO dataset, analyzed its access privileges, and measured the amount of data exposure. We chose depth over breadth due to the time-intensive nature of manual code and documentation analysis.

Since the dataset predated the widespread adoption of large language models, we focused on identifying rule-based chatbots by searching for recurring message patterns. This approach led us to discover the ``Discord Gophers Bot,'' which responds to messages prefixed with \texttt{?go} by providing tutorial links.\footnote{Although the chatbot's name was anonymized, consultation with the Gophers community confirmed it as the open-source project Discord Gophers Bot'' (\url{https://github.com/discord-gophers/dgobot}).} For instance, sending \texttt{?go tour} to the group prompts the chatbot to reply with ``A Tour of Go <https://tour.golang.org/welcome/1>,'' directing the user to the official Go tutorial. Similarly, the chatbot responds to \texttt{?go channels} by providing a YouTube tutorial link: ``Understanding channels <https://www.youtube.com/watch?v=KBZlN0izeiY>.''

\myparagraph{Overprivileged Bots.} To investigate whether a chatbot is overprivileged, we reviewed Discord Gophers Bot's source code and discovered a command handler snippet that processes messages starting with \texttt{?go}: \texttt{if !strings.HasPrefix(m.Content, "?go") \{ return \}}. Additionally, we found that the chatbot accesses the message sender's user ID and username through \texttt{Author.ID} and \texttt{Author.Username} attributes. Although the exact permissions granted to the chatbot are undocumented, this code analysis indicates that it requires broad permissions to read all messages and identify their senders, which is beyond its functional requirements.

\myparagraph{Disclosure of Irrelevant Messages.} 
Our analysis revealed that of 246,685 messages in the bot's group, only 580 (0.24\%) started with \texttt{?go} command. However, the bot had access to all messages---424 times more than necessary ($(246,685-580) / 580$). Among those messages the chatbot should not have access, we discovered several containing PII, including one case where a user accidentally shared his/her email address while pasting error logs. Such an incident exemplifies the privacy risks overprivileged chatbots pose in group settings. As of this writing, the bot remains active in the Gophers server, though a January 2022 update implemented Discord slash commands, restricting the bot's access to only pre-registered command messages. As Section~\ref{sec:Case Studies:official} will discuss, slash commands are now officially recommended due to their enhanced privacy and usability.

\myparagraph{Unnecessary Disclosure of User Metadata.} 
Recall that our code review revealed the chatbot's access to the sender's user ID and username.
However, the primary purpose of Discord Gophers Bot is to recommend tutorials, a function that does not require the sender's information. While the bot includes an admin feature to modify tutorial links, this feature only requires distinguishing admin from non-admin users, which is achievable through methods using pseudonyms. Moreover, our analysis of 246,685 messages revealed no instances of link modifications by admins, suggesting that the sender identification was unnecessary for all recorded conversations.

This case study shows how chatbots can be overprivileged by accessing unnecessary message content and metadata (RQ1-a). Moreover, when overprivileged, bots can accumulate large volumes of potentially sensitive information beyond their operational needs (RQ1-b).

\subsection{Case Study 2: Cross-Group Identification}
\label{sec:attack:cross-group}

\myparagraph{Methodology.}
When a user encounters the same chatbot in multiple groups where sender identities are exposed, the chatbot can link the user across these groups and aggregate inferred information. To quantify the probability of cross-group user identification (RQ2), we analyzed the Pushshift dataset~\cite{baumgartner2020pushshift}, focusing on Telegram chatbots with disabled privacy restrictions. Despite Telegram's default message access limitations, the chatbot metadata revealed that 718 (45.5\%) of 1,577 analyzed chatbots operated with the privacy mode disabled, granting them complete message access. We reconstructed a member list for each group by analyzing account metadata and messages.\footnote{We focused only on active members who sent messages, as the dataset does not include complete member lists.} We quantified user-chatbot relationships by counting how many groups each unique user-chatbot pair shared (i.e., encounter count).

\myparagraph{Prevalence of Cross-Group Chatbots.} Among the 718 chatbots studied, 253 (35.2\%) appeared in multiple groups, 44 (6\%) appeared in more than 10 groups, and three appeared in over 100 groups, significantly increasing the possibility of cross-group tracking. 

\myparagraph{User-Chatbot Encounters.} Among 1,168,344 users who joined at least one group, 42,508 (3.6\%) users encountered the same chatbot in multiple groups, with one user having repeated encounters with 134 different chatbots. Moreover, among the 4,155,927 user-chatbot pairs encountered at least once, 97,813 pairs (2.4\%) were encountered multiple times, and 96 pairs were encountered more than 10 times, with extreme cases of two pairs being encountered 55 times. These patterns demonstrate the extensive user data collection potential of an overprivileged chatbot.

%% file: sections/case_studies.tex
\section{Tension Between Mitigating Privacy Issues and Achieving E2EE}
\label{sec:Case Studies}

This section examines the privacy practices of popular group messaging platforms. Our survey shows that these messaging platforms face a fundamental tension between protecting user privacy from chatbots and enabling E2EE. While some platforms protect user privacy by filtering messages and hiding metadata, these protections are not trivially enforceable in the presence of E2EE, which prevents the inspection of message content. Our findings highlight the need for a practical solution to reconcile the two competing privacy goals.

\subsection{Methodology}
To understand the tension, we examined how popular messaging services address below privacy goals and analyzed their design decisions. 

\begin{itemize}
    \item \textbf{Mitigating Privacy Issues of Chatbots:} (1) Chatbots should only have access to necessary messages for their intended function. (2)  Chatbots should be unable to identify the sender of any message (e.g., through the permanent user identifier or profile information).
    \item \textbf{Achieving E2EE:} Messages should always be sent over E2EE channels that provide both FS and PCS.
\end{itemize}

We focused on globally popular platforms~\cite{Platform_usage_popularity} that support either E2EE or chatbot policies for analysis, including WhatsApp, Viber, Telegram, LINE, Discord, and Signal.
We also included Slack, which is mentioned in previous work on chatbot-related security issue~\cite{chen2022experimental}, and Keybase, which features cryptography-based restrictions on bot access control in groups~\cite{keybasebots}.

For platforms that officially support chatbots (\S\ref{sec:Case Studies:official}), we investigated chatbot permissions by creating chatbots using the official APIs and following the respective platform guidelines. 
On each platform, we set up a group, added a chatbot, and tested its behavior by sending messages in various formats. These included standard messages, messages that mentioned or tagged the chatbot, and messages that use predefined formats (e.g., commands) on platforms that support such functionalities. We then used the chatbot APIs to analyze the messages the chatbot received and the metadata included in these messages. We also investigated whether message contents were encrypted using E2EE by using Wireshark to capture and analyze network packets.

For platforms without official chatbot support (\S\ref{sec:Case Studies:unofficial}), we searched for well-known unofficial open source libraries that claim to enable chatbot functionality, and analyzed their code. We chose not to create chatbots using these libraries because the programmatically-simulated user’s E2EE status and message access policies are predictably identical to those of regular users. Also, using these libraries would violate the platforms’ terms of service.

\subsection{Platforms with Chatbot Support}
\label{sec:Case Studies:official}

Most platforms that support chatbots in group chats enforce policies to regulate their access control.
In the following discussion, we examine the design measures these platforms employ to prevent chatbots from learning excessive information, and evaluate whether they maintain E2EE in the presence of chatbots.

\myparagraph{Telegram.}
Telegram restricts message access for chatbots, but still exposes the sender's identity in messages and does not supports E2EE.
By default, chatbots in a group operate in \textit{Privacy Mode}, allowing them to access only system messages or messages that mention them using pre-registered commands~\cite{telegram2023privacy2}.
However, they can still access metadata about the sender~\cite{telegram2023api}. 
Additionally, Telegram’s MTProto 2.0 protocol~\cite{MTProto} only implements group chats using server-client encryption rather than E2EE~\cite{Telegramcloudchats}.

\myparagraph{Slack.}
Slack provides two different permission models for chatbots: the granular bot permission scopes and the legacy bot permission scopes~\cite{slack2023permission}. The granular bot permission scopes are designed to improve privacy by implementing a new scope that limits chatbots to messages in which they are specifically mentioned. When mentioned, the chatbot can identify the sender within the group using a unique identifier, but this identifier does not directly link to the sender's profile, thereby achieving pseudonymity against chatbots. In contrast, chatbots under the legacy bot permission scopes have unrestricted access to messages.
Slack encrypts all messages and data at rest and in transit.
However, it does not provide E2EE, which means that messages can be decrypted by Slack on their servers~\cite{slack_whitepaper}.

\myparagraph{Discord.}
Discord employs a fine-grained permission model for chatbots, covering message access in groups. By default, chatbots cannot read messages unless explicitly mentioned. Discord recommends using the new \textit{Interactions} API~\cite{discord_interactions}, where chatbots can interact with messages with slash commands, buttons, or menus.
Additionally, chatbot developers can request permission to access all messages, but this requires manual verification and approval, including checking the presence of a privacy policy, as outlined in Discord’s criteria for approval~\cite{discord_review_policy}.
On top of that, Discord does not hide the sender identities from the chatbots and does not support E2EE.

\myparagraph{LINE.}
Group chats involving chatbots in LINE are not end-to-end encrypted, and privacy issues from chatbots are not mitigated. LINE offers group E2EE through a mechanism called ``Letter Sealing,'' which is similar to the Sender Keys Protocol and provides FS~\cite{line2021encryption, LINEtransparency}. However, Letter Sealing is automatically disabled in groups when a chatbot is added. Furthermore, LINE does not implement any specific measures to reduce the privacy risks associated with chatbots, which have the same access privileges as regular users, including full access to messages and sender identities~\cite{LINEprotocol}.

\myparagraph{Keybase.}  
All messages shared in Keybase Chat are transmitted through channels with E2EE, and it allows certain messages to be revealed to the chatbot while maintaining E2EE.
Keybase~\cite{keybasebasic} uses a strategy similar to Telegram's privacy mode to restrict the chatbot's access to messages.
In addition to messages that explicitly mention the chatbot using \texttt{@Bot}, Keybase also shares messages that match commands pre-registered by the chatbot, offering flexibility. 
Message metadata reveals the sender's username, allowing the chatbot to learn the sender's identity.

For messages intended for the chatbot, Keybase uses bot-specific keys shared between the group members and the chatbot to encrypt the messages.
The bot-specific key is derived from the team key, a key shared by all group members except the chatbots, and is then sent to the chatbot, encrypted with the chatbot's user key. 
By encrypting with the bot-specific key, both the chatbot and the group members can decrypt the message.
Messages not intended for the chatbot are encrypted using the team key. As a result, the chatbot is aware of these messages but cannot decrypt the content since it lacks access to the team key~\cite{keybasebots,keybasechat}.
Due to its similar design to the Sender Keys protocol, it likely does not support PCS. 
This claim is supported by the absence of PCS in their documentation~\cite{keybasechat} or security assessment reports~\cite{keybasesecurity}, and the lack of key renegotiation features in the current design.
Furthermore, because group members are responsible for initiating the creation of bot-specific keys, chatbots cannot initiate interactions or update keys, even if the keys are compromised.

\subsection{Platforms Lacking Chatbot Support} 
\label{sec:Case Studies:unofficial}
Platforms like WhatsApp, Viber, and Signal do not natively support chatbot functionality in group chats. As a workaround, some developers create automated users to simulate chatbots on these platforms. This has been confirmed by reviewing the source code of third-party chatbots~\cite{3rdparty_whatsapp_bot, whatsapp_webjs}.
Since these unofficial chatbots operate as regular users, they have the same permissions, including full access to all group messages and sender identities. %
While WhatsApp, Viber, and Signal offer E2EE in group chats, as outlined in their respective documentation~\cite{WhatsApp, Viber_Whitepaper, libsignal}, most of these platforms use a Sender Keys-like protocol, which does not support PCS. Only Signal’s private groups~\cite{privategroup}, which use pairwise encryption for small groups, support PCS.

\subsection{Findings}
\label{ssec:findings}

Our survey revealed two common approaches among popular group messaging platforms, highlighting the tension between mitigating overprivileged chatbots and achieving E2EE.

Most platforms, such as Telegram, Discord, and Slack, forgo E2EE when interacting with chatbots, allowing service providers to control what information should be shared with chatbots. In contrast, some platforms, such as WhatsApp, Viber, and Signal, focus on providing E2EE for group chats and do not support chatbots in group chats. However, the development of unregulated third-party bots, while retaining full E2EE capabilities, may have access to all messages. A notable exception to these two approaches is Keybase. It allows E2EE and message filtering to coexist, but lacks some essential security features found in modern E2EE protocols. 
In addition, none of these group messaging platforms effectively address the privacy concerns associated with unnecessarily revealing sender identities. Table~\ref{tab:messaging_platforms} summarizes each platform's support for the desired privacy goals. 

Our findings suggest that a novel technique is needed to resolve the tension between mitigating overprivileged chatbots and achieving E2EE. In the rest of the paper, we will present our proposed protocol to resolve this tension.

\newcolumntype{R}[2]{%
    >{\adjustbox{angle=#1,lap=\width-(#2)}\bgroup}%
    l%
    <{\egroup}%
}
\newcommand*\rot{\multicolumn{1}{R{60}{1em}}}%

\begin{table}[htbp]
\centering
\begin{footnotesize}
\begin{tabular}{c c c c c c c c c}
\hline  
& \rot{WhatsApp*} & \rot{Viber*} & \rot{Signal*} & \rot{LINE} & \rot{Telegram} & \rot{Discord} & \rot{Slack} & \rot{Keybase} \\
\hline
Group E2EE & \CIRCLE & \CIRCLE & \CIRCLE & \CIRCLE & - & - & - & \CIRCLE \\
\hline
\makecell{Group E2EE (w\textbackslash~bots)} & \CIRCLE & \CIRCLE & \CIRCLE & - & - & - & - & \CIRCLE \\
\hline
\makecell{PCS} & - & - & \LEFTcircle & - & - & - & - & - \\
\hline
\makecell{Message access control} & - & - & - & - & \CIRCLE & \CIRCLE & \CIRCLE & \CIRCLE \\
\hline
Hide sender & - & - & - & - & - & - & \CIRCLE & - \\
\hline
\end{tabular}
\end{footnotesize}
\vspace{1pt}

\CIRCLE: fully support, \LEFTcircle: partially support, -: no support
\caption{Comparison of security properties on popular messaging platforms. 
Signal partially supports PCS only in private groups.
Platforms marked with * do not officially support chatbots.
} 
\label{tab:messaging_platforms}
\end{table}

%% file: sections/protocol.tex
\section{\cmrt: A Privacy-Preserving Secure Group Messaging Protocol}\label{sec:cmrt}

We aim to design a secure group messaging protocol that addresses two privacy concerns posed by overprivileged chatbots: (1) lack of message access control, and (2) unnecessary disclosure of sender identities, all while ensuring robust E2EE with FS and PCS. 
Our solution, \cmrt, builds on a tree-based continuous group key agreement (CGKA) scheme, enhancing the traditional key tree structure to overcome its limitations and support our desired properties. This design enables selective message access and sender anonymity. Additionally, we introduce extensions to support pseudonymity and enhance user privacy by hiding message triggers from service providers.

\subsection{Desired Properties and Challenges}
\label{design:properties-and-challenges}

We first outline the desired properties of our messaging protocol: \textit{selective message access} and \textit{sender anonymity} to address the two privacy issues, and challenges achieving them using existing protocols.

\myparagraph{Selective Message Access.} 
The concept of selective message access refers to the ability to restrict chatbots from accessing messages that are irrelevant to their functionality. Implementing this poses several challenges. First, E2EE makes it impractical for service providers to filter messages for chatbots, as platforms like Telegram and Discord currently do, since the providers cannot read the encrypted messages. Second, to maintain confidentiality, chatbots should not have access to the encryption keys used for messages not intended for them, as possessing the ciphertext alone would allow them to decrypt the messages if they had the necessary keys.

However, existing group messaging protocols such as Sender Keys do not facilitate selective key access, as they use static keys that do not change per message. 
Similarly, CGKA-based protocols that treat chatbots as users require all participants, including chatbots, to catch up with all key updates, also preventing selective access.
Keybase's protocol allows selective key access by having a chatbot-specific sender key that is shared by all group members, but does not allow chatbots to update their keys and does not achieve PCS. 
Therefore, there is a need for a group messaging protocol that performs message filtering on the client side and implements a key management strategy that allows selective key access for chatbots, ensuring that they can only decrypt messages meant for them.

\myparagraph{Sender Anonymity.} 
Sender anonymity aims to hide the sender’s identity from chatbots. %
Chen et al. define a form of sender anonymity called internal group anonymity (IGA), which guarantees sender indistinguishability among all group members~\cite{chen2020anonymous}. However, our scenario requires a directional version: while users should be indistinguishable to chatbots, users should still be able to distinguish chatbots. To address this, we introduce a new notion called selective IGA, which provides directional sender anonymity. We also aim to achieve pseudonymity, another notion of sender anonymity that allows chatbots to differentiate between users without knowing their real identities, as in the case of anti-spam bots.

Despite this, no existing group messaging protocol provides the required level of sender anonymity. In the Sender Keys protocol, each sender can be uniquely identified through their individual sender keys. Similarly, in tree-based CGKA schemes like ART and TreeKEM, the ``update path''~\cite{rfc9420} reveals node information from the sender’s node to the root, which can be used to identify the sender. To address this, Chen et al. proposed Anonymous ART (AART)~\cite{chen2020anonymous}, which supports IGA, but does not support selective message access for the same reasons as other CGKA schemes.

\subsection{Design Overview}~\label{cmrt:overview}
\begin{figure*}[t]
    \centering
    \begin{subfigure}[t]{0.225\textwidth}
         \centering
         \includegraphics[width=\linewidth]{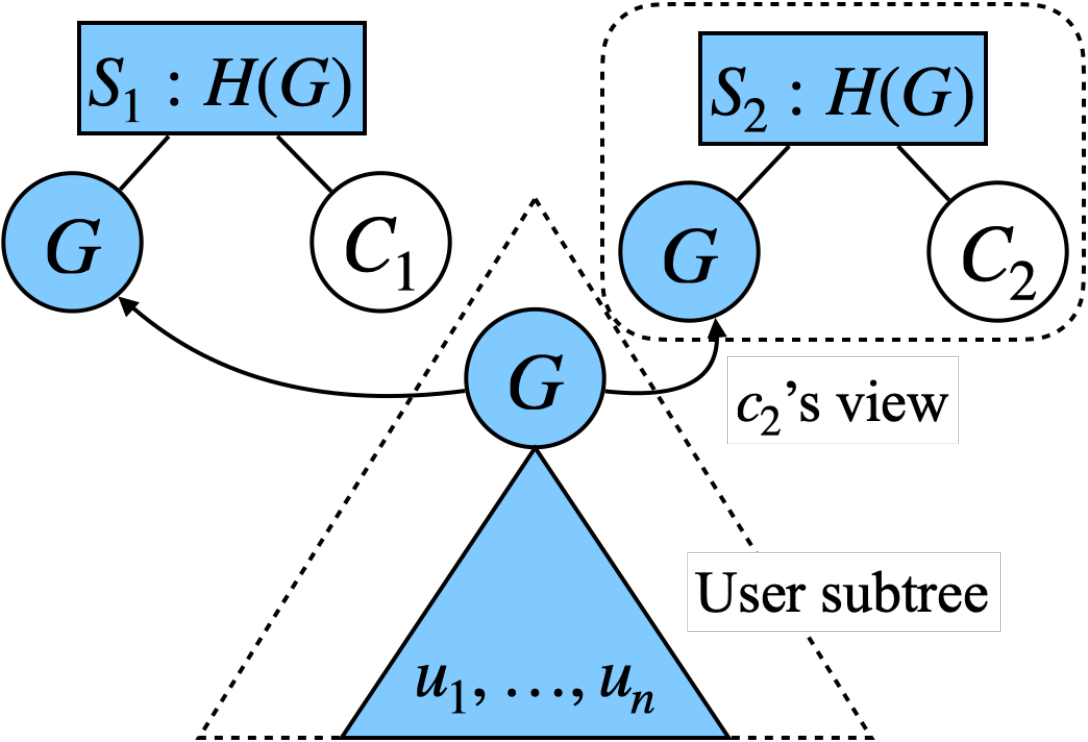}
         \caption{Users perform an update to obtain the group secret $G$ and update the shared secrets for $C_1, C_2$. Secrets updated in this phase are colored blue.}
     \end{subfigure}
     \hfill
     \begin{subfigure}[t]{0.22\textwidth}
         \centering
         \includegraphics[width=\linewidth]{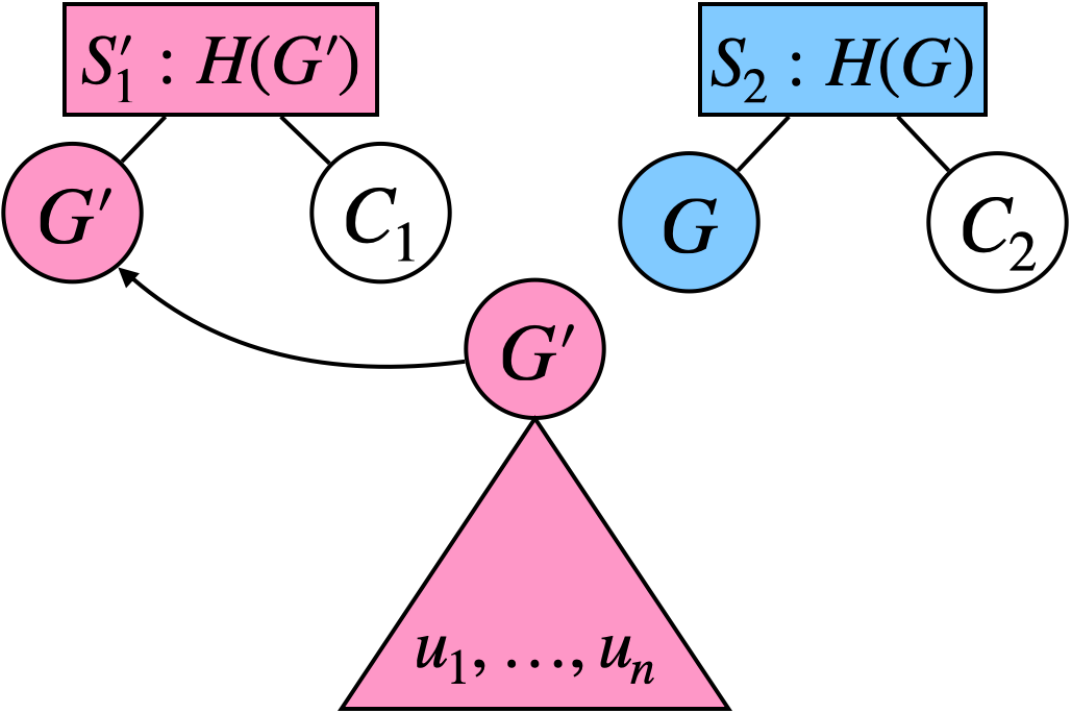}
         \caption{Users perform an update to obtain the group secret $G'$ and update the shared secrets for $C_1$. Secrets updated in this phase are colored red.}
     \end{subfigure}
     \hfill
     \begin{subfigure}[t]{0.22\textwidth}
         \centering
         \includegraphics[width=\linewidth]{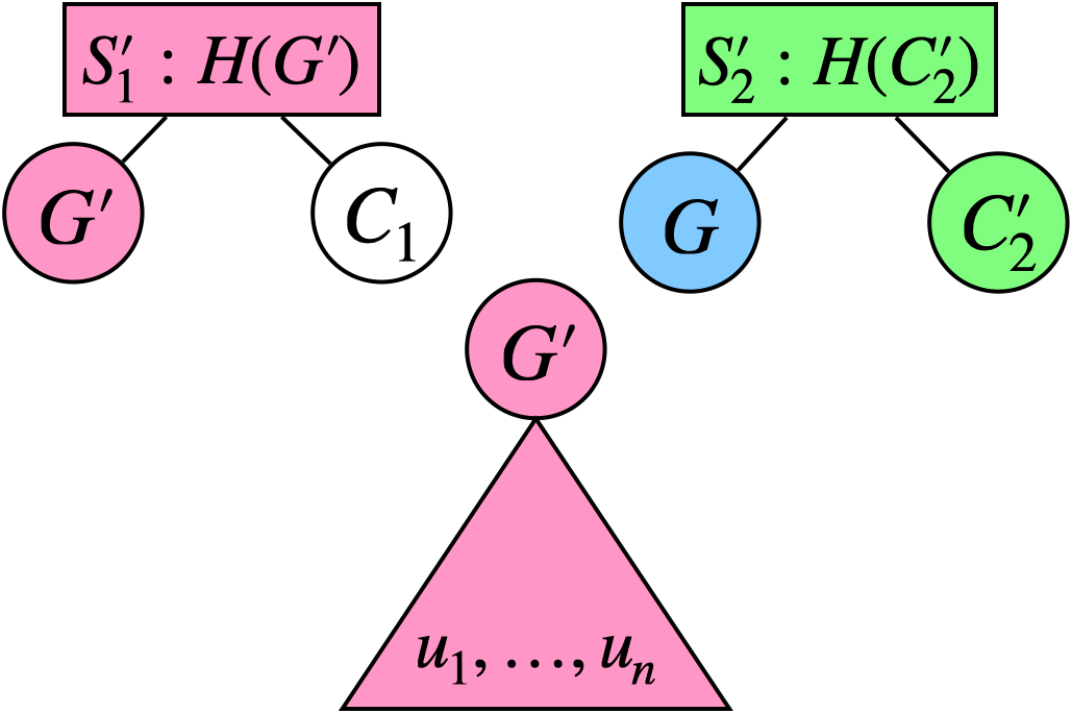}
         \caption{$C_2$ performs an update. Secrets updated during this phase are colored green. Updates initiated by chatbots will not trigger an update for the user subtree.}
         \label{fig:y equals x}
     \end{subfigure}
     \hfill
     \begin{subfigure}[t]{0.22\textwidth}
         \centering
         \includegraphics[width=\linewidth]{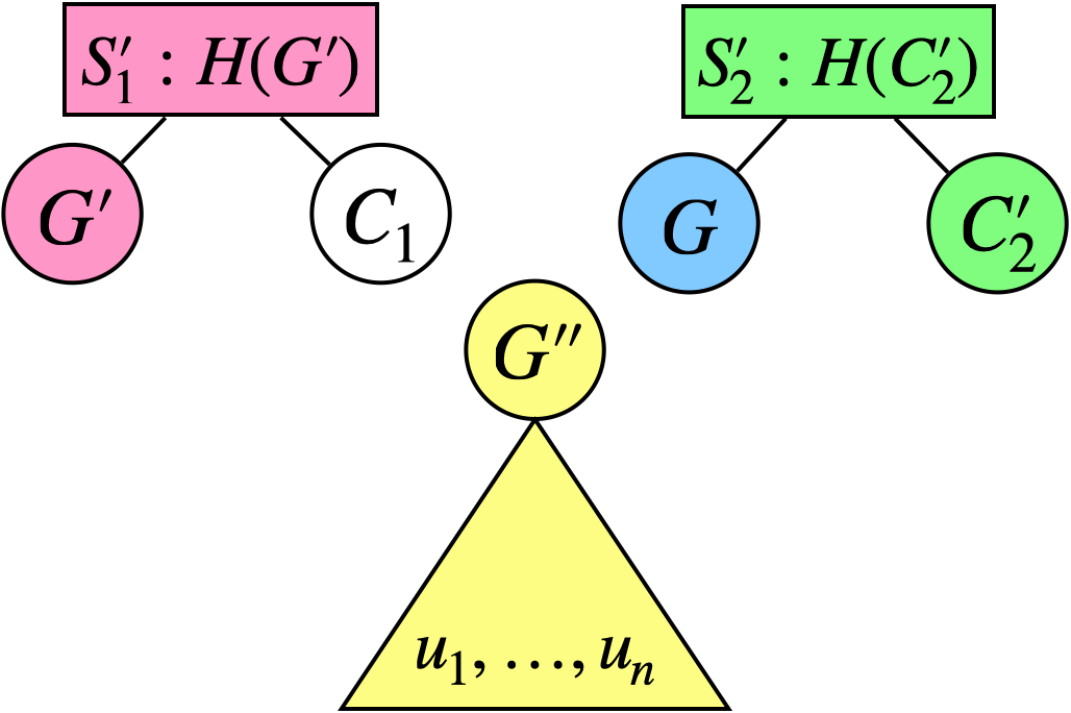}
         \caption{Users trigger an update only for user subtree, resulting the group secret $G''$. Secrets updated during this phase are colored yellow.}
     \end{subfigure}
    \caption{Illustration of \cmrtt with users $u_1, \dots, u_n$ and chatbots $c_1, c_2$. Users share the group secret $G$ from the \textit{user subtrees} (triangles), while $C_1, C_2$ are secrets for chatbots $c_1, c_2$, respectively. The rectangles represent secrets shared between the group and each chatbot. The arrows indicate secret assignments, and the lines indicate parent-child relationships, where a child knows the secret of its parent. For example, in (b), $G'$ represents the group secret, while $S_1'$ is the secret shared between $u_1, \dots, u_n$ and $C_1$. In (c), the chatbot can send the fresh key $S_2'$ to the users using the public key derived from $G$.
    }
    \label{fig:cmrt_overview}
\end{figure*}

Our solution is built on a tree-based CGKA scheme, as it is the most common key agreement scheme that achieves both FS and PCS. Before detailing the specific components of our protocol, we first explain how our approach addresses the challenges outlined in Section~\ref{design:properties-and-challenges}. The core of our approach lies in modifying the structure of key tree used in CGKA. We concatenate the tree root of the CGKA with several new ``root nodes,'' each representing a separate group key.

\myparagraph{Selective Message Access.} 
One of the main challenge achieving selective message access is that chatbots may share inconsistent keys with users, since if chatbots do not receive all messages, they may miss key updates.
However, chatbots using traditional CGKA require all key update messages to maintain key consistency. 
To solve this, our construction employs multiple root nodes, each corresponding to a separate group key—one for each chatbot. Each chatbot maintains its own key state shared with the user group, and its key is updated only when a message is specifically intended for that chatbot (as illustrated in Figure~\ref{fig:cmrt_overview}). If no key update occurs for a given chatbot, its key remains unchanged, allowing it to continue encrypting messages with the current key.

\myparagraph{Selective IGA.} Tree-based CGKA protocols typically fail to provide sender anonymity because the sender’s identity can be inferred from the \textit{update path}—the path from the sender’s node to the root. To mitigate this, we modify the structure of key tree by separating group members into two subtrees: a \textit{user subtree} containing all user nodes, and a \textit{chatbot subtree} containing only the chatbot nodes (as shown in Figure~\ref{fig:cmrt_overview}).
During a user-initiated key update, the chatbot is only made aware of the root of the user subtree, rather than the full details of the individual nodes within the subtree. This design allows the chatbot to compute a shared secret with the entire group without knowing which specific user initiated the update, thereby preserving the sender’s anonymity.

The resulting structure of key tree, called the Compressed Multi-Roots Tree (\cmrtt), achieves these two additional security properties by assigning each chatbot a dedicated subtree while sharing a root node with the user subtree. From the users' perspective, their tree connects to multiple root nodes, each corresponding to a different chatbot. In contrast, each chatbot perceives itself as part of a smaller 3-node tree that links only to the root of the user subtree. Key updates are managed efficiently, with users only storing the root nodes, optimizing storage while maintaining asynchronous states across different chatbots, as shown in Figure~\ref{fig:cmrt_overview}. This ``multi-root'' structure is ``compressed'' to balance data storage efficiency with strong security guarantees.

\subsection{Building Blocks}~\label{cmrt:buliding-blocks}

This subsection presents the building blocks of our construction, including cryptographic primitives, a formal definition of CGKA, and TreeKEM, the CGKA construction that serves as the foundation for our solution. Throughout this section, $\sample$ denotes assignment from a randomized algorithm, while $\gets$ represents assignment from a deterministic one.

\myparagraph{Cryptographic Primitives.}
The public-key encryption (PKE) scheme $\pke = (\pkeg, \pkenc, \pkdec)$ consists of key generation $(\sk, \pk) \sample \pkeg(\secparam)$, encryption $c \sample \pkenc(\pk, m)$ using the public key $\pk$, and decryption $m \gets \pkdec(\sk, c)$ that retrieves the original message $m$ from the ciphertext $c$. 
We also use a digital signature scheme $\sig = (\sign, \verify)$, where $s \sample \sign(\sk, m)$ signs a message $m$, and $\verify(\pk, s, m)$ verifies it using the corresponding public key. 
Additionally, a collision-resistant hash function $\hash: \bin^\secpar \rightarrow \bin^\secpar$ is used.

\myparagraph{Continuous Group Key Agreement (CGKA).}
As detailed in Alwen et al.~\cite{alwen2020security}, a CGKA scheme comprises operations for key agreement in secure group communication, enabling all members to share a common secret key. This scheme allows for dynamic adjustments in group membership, including adding or removing members. The shared key is updated whenever there are changes to the membership or upon request by any group member. For instance, the key can be refreshed with every sent message, which helps guarantee that future key compromises do not impact the security of previously sent messages. The syntax of CGKA is defined as follows, as outlined by Alwen et al.~\cite{alwen2020security}.

\begin{definition}\label{def-cgka}
    A CGKA scheme is a tuple of the following algorithms $\cgka=(\init, \creategroup, \add, \rem, \upd, \proc)$:
    \begin{itemize}
        \item $\gamma \sample \init(\mathsf{ID})$ takes a user ID $\mathsf{ID}$ and outputs an initial state $\gamma$.    
        \item $(\gamma^{\,\,\prime}, T) \sample \creategroup(\gamma, \mathsf{ID_1}, \dots, \mathsf{ID_n)}$ takes a state $\gamma$ and a list of user IDs $\mathsf{ID_1}, \dots, \mathsf{ID_n}$. This outputs a new state $\gamma^{\,\,\prime}$ and control message $T$.  
        \item $(\gamma^{\,\,\prime}, T) \sample \mathsf{add}(\gamma, \mathsf{ID})$ takes a state $\gamma$ and a user $\mathsf{ID}$ to add, and outputs a new state $\gamma^{\,\,\prime}$ and control message $T$.    
        \item $(\gamma^{\,\,\prime},T) \sample \mathsf{rem}(\gamma, \mathsf{ID})$ takes a state $\gamma$ and a user $\mathsf{ID}$ to remove, and outputs a new state $\gamma^{\,\,\prime}$ and control message $T$.
        \item $(\gamma^{\,\,\prime},T) \sample \mathsf{upd}(\gamma)$ takes a state $\gamma$ and outputs a new state $\gamma^{\,\,\prime}$ and control message $T$.
        \item $(\gamma^{\,\,\prime}, k) \gets \mathsf{proc}(\gamma, T)$ takes a state $\gamma$ and a control message $T$ or $W$. This outputs an updated state $\gamma^{\,\,\prime}$ and a fresh group key $k$.
    \end{itemize}
\end{definition}

To initiate the CGKA scheme, each user, identified by $\id$, starts by initializing their state with $\init(\id)$. Upon initialized, a user can create a group using $\creategroup$ with the initial members' identifiers. Subsequent group modifications, such as adding members with $\add$, removing them via $\rem$, or updating personal key material with $\upd$, each triggers an update to the group key and generates a control message $T$. Group members keep their states synchronized by processing these control messages via $\proc$, ensuring they all share the same group key $k$.

\myparagraph{TreeKEM.}
TreeKEM~\cite{bhargavan2018treekem} is a tree-based CGKA scheme constructed with a hashing key tree. Each node in this tree holds a secret accessible only to the members within its subtree, and each member is assigned to a leaf node. The secret of the root node serves as the shared secret for the entire group.
In TreeKEM, let $S_i \in \bin^\secpar$ denote a member's $i$-th secret from the leaf. The secret $S_i$ is computed as the hash of the secret $S_{i-1}$ from one of its child nodes, specifically the last child that updates the secret. Additionally, each node contains a pair of public-private keys $(sk_i, pk_i)$ generated from its secret $S_i$ using $\mathsf{PKEG}$. The node information can be computed by $(sk_i, pk_i) \leftarrow \mathsf{PKEG}(S_i)$ where $S_i \leftarrow H(S_{i-1})$.

To perform key update, a member (associated with one of the leaf node) randomly generates a new secret and iteratively computes the secrets along the path to the root node via hashing. The member then notifies other members of the new secrets by encrypting them with the public keys of the sibling nodes. Specifically, for a node $v$ with a new secret $S'$, the member encrypts $S'$ using the public key of node $\mathsf{sibling}(v)$ and sends the ciphertext to the members under $\mathsf{sibling}(v)$, where $\mathsf{sibling}(v)$ denotes the sibling node of $v$. The member also publishes all new public keys along the updated path.

\subsection{\cmrt Operations}
\label{sec:protocol:basesline}

We now introduce the core operations of our secure group messaging protocol, \cmrt. \cmrt extends existing secure group messaging protocols by adding chatbot-specific operations: chatbot addition/removal, message sending by chatbots, and trigger functions. It uses $\cmrtt$, which relies on an existing tree-based CGKA scheme for user subtree management, to enable selective message access and selective IGA. 
While \cmrt handles communication between users and chatbots, interactions among users continue to rely on the underlying secure group messaging protocols.
We assume that the service provider acts as a public key infrastructure (PKI) where chatbots register their public keys and signed trigger functions.
The pseudocode for the baseline protocol is presented in Appendix~\ref{appendix:pseudo}.

\paragraph{Trigger Function.} To enable client-side message filtering, we introduce the concept of \textit{trigger function} represented as $f_{\cid}: \mathcal{M} \rightarrow \{0, 1\}$, where $\mathcal{M}$ denotes the message space and $\cid$ denotes the unique identifier of a chatbot. The function returns $1$ if a message is considered relevant to a chatbot identified by $\cid$, and $0$ otherwise. The function is evaluated on user's device before a message is sent. The function can be implemented in various form, such as a snippet code executed in a sandbox environment, or a checkbox that asks for permission. 

\myparagraph{State Initialization.} The $\cmrtt$ state for each entity, including a user or a chatbot, is denoted as $\gamma$. For users, the $\cmrtt$ state includes the state of the underlying CGKA scheme $\gamma.s0$ and a dictionary $\gamma.\chatbots$ indexed by chatbot identifier to record leaf nodes information for each chatbot in a group. For chatbots, the $\cmrtt$ state consists of its identifier $\gamma.\me$, the root public key of the user subtree $\gamma.\gpk$, a long-term PKE key pair $(\gamma.\sk_\cid, \gamma.\pk_\cid)$ for identity authentication, and a PKE key pair $(\gamma.\csk, \gamma.\cpk)$ associated with the chatbot's tree node.

When a chatbot identified by $\cid$ is initialized, it registers the long-term public key $\pk_\cid$, the trigger function $f_\cid$ and its signature $s \gets \sign(\sk_\cid, f_\cid)$ to the service provider.

\myparagraph{Group Creation.} We assume only a user can create a group, and a group initially contains only users. The process of creating a group with identities $\id_1, \dots, \id_n$ involves the following steps: (1) A user with state $\gamma$ initiates the group using the CGKA scheme's create algorithm: $(\gamma.s0, T) \gets \cgka.\creategroup(\gamma, \id_1, \dots, \id_n)$. (2) This user then broadcasts the control message $T$ to the users identified by $\id_1, \dots, \id_n$. (3) Upon receiving the message, each user processes the control message $T: (\gamma.s0, k) \gets \proc(T)$.

\myparagraph{Adding or Removing Users.} Similar to group creation, a user with state $\gamma$ calls the CGKA scheme using $(\gamma.s0, T) \gets \cgka.\add(\id')$ or $(\gamma.s0, T) \gets \cgka.\rem(\id')$, where $\id'$ is the identifier of the user to be added or removed. After that, the user broadcasts the control message $T$ to all user members, which is then processed by $(\gamma.s0, k) \gets \proc(T)$, resulting in a fresh group key $k$.

\myparagraph{Adding a Chatbot.} Adding a chatbot identified by a chatbot ID $\cid$ to a group involves four steps: 
(1) The initiating user retrieves chatbot's public key $\pk_\cid$, samples a leaf secret $k$, computes chatbot subtree's public key $\cpk$ by $(\csk, \cpk) \gets \pkeg(k)$, and encrypts the secret: $e \gets \pkenc(\pk_\cid, k)$. (2) The user broadcasts root public key of user's subtree $\gamma.s0.\gpk$, the encrypted secret $e$, and public key $\cpk$ of chatbot's subtree, to all group members and the chatbot.
(3) All users retrieve the chatbot's public key $\pk_\cid$, trigger function $f_\cid$, and the associated signature $s$ from the service provider and store the chatbot's information into their respective states: $\gamma.\chatbots[\cid] \gets (f_\cid, \gamma.s0.\gsk, \cpk)$ if the signature is successfully verified by $\verify(\pk_\cid, s, f_\cid)$.
(4) The chatbot decrypts the initial secret $k \gets \pkdec(\gamma.\sk_\cid, e)$, and stores the received root public key $\gpk$, and the derived PKE key pair $(\csk, \cpk) \gets \pkeg(k)$. 
This process initializes a TreeKEM key tree between the user group and the chatbot.

\myparagraph{Removing a Chatbot.} Removing a chatbot identified by a chatbot ID $\cid$ involves the following steps. (1) The initiating user broadcasts the decision to remove the chatbot to all group members. (2) All users then clear the corresponding entry from their records: $\gamma.\chatbots[\cid] \gets \perp$. (3) The removed chatbot also clears its information about the user group: $\gamma.\gpk \gets \perp$.

\myparagraph{Users Sending a Message to Chatbot.} When a user with state $\gamma$ sends a message to chatbots, the process involves initiating an underlying CGKA (user subtree) key update (step 2) and a TreeKEM update (step 2 and 4), and encrypting the message using the newly derived shared secret (step 3). The steps are as follows: (1) Obtains a fresh group key $k$ from the CGKA scheme: $(\gamma.s0, T_0) \gets \cgka.\upd(\gamma.s0); (\gamma.s0, k) \gets \proc(\gamma.s0, T_0)$. (2) Computes the group PKE key pair $(\gsk, \gpk) \gets \pkeg(k)$ and a new message encryption key $k' \gets \hash(k)$. (3) Encrypts the message $m$: $c \gets \enc(k', m)$ and prepares the control message $T=(T_0, c, \gpk)$. (4) For each chatbot identified by $\cid$ that requires the message (i.e., $f_\cid(m) = 1$), retrieves the chatbot's public key $\cpk \gets \gamma.\chatbots[\cid].\cpk$, appends $(\cid, \pkenc(\cpk, k'))$ to the control message $T$, and updates the group secret key for that chatbot $\gamma.\chatbots[\cid].\gsk \gets \gsk$.  (5) Finally, broadcasts the control message $T$ to the group, including the chatbots. 

We require that either the CGKA's control message $T_0$ is removed from the control message $T$ before forwarded to the chatbots, or $T_0$ is encrypted in a way similar to MLS's Private Message~\cite{rfc9420}, so that chatbots cannot decrypt $T_0$ nor learn the sender's identity. When sending a message to both users and chatbots, this process operates in parallel with the original secure group messaging protocol, which handles ciphertext generation and key updates for users.

A user with state $\gamma$ receiving the control message $T$ updates the keys using the following steps: (1) Computes the fresh group key through: $(\gamma.s0, k) \gets \cgka.\proc(T_0)$. (2) Computes the group PKE key pair $(\gsk, \gpk) \gets \pkeg(k)$.
(3) For each $\cid$ in the control message, updates the group secret key $\gamma.\chatbots[\cid].\gsk \gets \gsk$.

A chatbot with state $\gamma$ receiving the control message $T$ first checks if its identifier $\cid$ is included. If not, the message is deemed unrelated to this bot. If the identifier is found, the chatbot proceeds to decrypt the message using the following steps: (1) Decrypts the chatbot encryption key $k'$ from the ciphertext $e$ associated with $\cid$: $k' \gets \pkdec(\gamma.\csk, e)$. (2) Decrypts the message content $m$ using $m \gets \dec(k', c)$. (3) Updates the group public key in its state to $\gamma.\gpk \gets \gpk$.

\myparagraph{Chatbot Sending a Message.} When a chatbot with state $\gamma$ and identified by $\cid$ sends a message to users, it issues a TreeKEM update and encrypts the message using the newly derived shared secret. 
The steps are as follows: (1) Randomly generates a key: $k \sample \bin^\secpar$. (2) Computes the message encryption key $k' \gets \hash(k)$ and the PKE key pair $(\csk, \cpk) \gets \pkeg(k)$ for chatbot's tree node. (3) Encrypts the message $m$: $c \gets \enc(k', m)$ and encrypts the encryption key using group public key: $e \gets \pkenc(\gamma.\gpk, k')$ (4) Updates the chatbot private key $\gamma.\csk \gets \csk$. (5) Finally, broadcasts the entire control message $T=(\cid, c, e, \cpk)$ to the group.

A user with state $\gamma$ receiving the control message $T$ decrypts the message using the following steps: (1) Decrypts the encryption key using the dedicated group private key for $\cid$: $k' \gets \pkdec(\gamma.\chatbots[\cid].\gsk, e)$. (2) Decrypts the message content: $m \gets \dec(k', c)$. (3)  Updates the chatbot public key $\gamma.\chatbots[\cid].\cpk \gets \cpk$.

\subsection{\cmrt Extensions}

We present two \cmrt extensions: pseudonym visibility for stateful chatbot applications and trigger concealment from service providers for enhanced user privacy.

\paragraph{Pseudonymity}
Achieving pseudonymity involves additional two steps: (1) The group member registers a pseudonym anonymously
; (2) When a group member sends a message using the pseudonym, the chatbot verifies that the message truly comes from the authenticated sender behind the pseudonym. Our protocol should ensure that in both steps, the chatbot remains unaware of the exact identity of the member.

In our protocol, a pseudonym contains a PKE key pair used as message signing key. 
In the first step, a member generates an ephemeral identity key pair $(sk_e, pk_e)$ and registers the identity by broadcasting the public key $pk_e$ to all chatbots using the baseline protocol, then each chatbot records $pk_e$ as a new ephemeral identity in the group. 
In the second step, the member signs the message using $sk_e$, and the chatbot verifies the signature with $pk_e$ to check the legitimacy of the message sender. Because baseline protocol provides sender anonymity, the chatbot cannot associate two pseudonyms with the same user. The users can easily change their pseudonyms by registering new ones.

\paragraph{Trigger Concealment from Service Providers}
Chatbots are typically triggered by messages matching specific patterns, which leaks information about message content to service providers. For example, if a phishing detection bot is triggered, the provider might deduce the message contains a URL.
To enhance user privacy, senders can transmit control messages to chatbots without including key updates for those that should not be triggered. In step (4) of the procedure for sending a message to a chatbot, described in Section~\ref{sec:protocol:basesline}, when $f_\cid(m)$ is false, meaning that the chatbot identified by $\cid$ should not receive the message, the sender appends $(\cid, r)$, where $r$ represents the random bytes that should be indistinguishable from the actual ciphertext of the key update, thereby ``hiding'' the triggering event. These chatbots would not be able to decrypt the new secret and would therefore drop the message.

\subsection{Integration with Existing Secure Group Messaging Protocols} 
\label{sec:protocol:integration}
\cmrt needs to run alongside a secure group messaging protocol. There are two strategies for integration, depending on whether the underlying secure group messaging protocol uses a CGKA key tree. 

\myparagraph{Protocols without CGKA.} 
Our protocol can coexist with secure group messaging protocols that do not rely on CGKA, such as the Sender Keys Protocol, without interfering with their operations. When creating a group and updating members, in addition to the original key exchange procedures, such as sharing sender keys, users also maintain a complete $\cmrtt$, including the user's subtree. When sending a message, the sender follows the original procedure to communicate with other users, which, in the case of the Sender Keys Protocol, involves encrypting messages with their sender key, and follows \cmrt to send messages to chatbots by encrypting with each chatbot's root key and issuing key updates. 

\myparagraph{Protocols with CGKA.} The advantages of our protocol can be highlighted when integrated with secure group messaging protocols that use tree-based CGKA, such as MLS. 
In these cases, the protocol's key management already relies on a tree-based CGKA key tree to generate a shared secret for group members, and $\cmrtt$ can utilize the existing tree as the user's subtree. This further reduces the overhead of maintaining the \cmrt, since the effort of storing and maintaining the user's subtree is already included in these protocols.

%% file: sections/evaluation.tex
\section{Evaluation}

\subsection{Security Analysis}

\begin{table}[htbp]
    \centering
    \begin{small}
    \begin{tabular}{c|c c|c|c}
        \hline
         & \multicolumn{2}{c|}{E2EE} & Sender & \multirow{2}{*}{SMA}  \\ \cline{2-3}
         & FS & PCS & Anonymity &   \\
         \hline
         Sender Keys & \CIRCLE & - & - & - \\
         \hline
         Keybase & \CIRCLE & - & - & \CIRCLE \\
         \hline
         MLS (TreeKEM) & \CIRCLE & \CIRCLE & - & - \\
         \hline
         AART~\cite{chen2020anonymous} & \CIRCLE & \CIRCLE & IGA & - \\
         \hline
         \cmrt & \CIRCLE & \CIRCLE & Selective IGA & \CIRCLE \\
         \hline
    \end{tabular}
    \end{small}
    \caption{Security comparisons between 
    secure group messaging protocols. SMA stands for Selective Message Access.}
    \label{table:security_comparisons}
\end{table}

This section provides a security analysis to show that our protocol satisfies the desired security properties, while achieving FS and PCS, as shown in Table~\ref{table:security_comparisons}. The analysis is based on the assumption that the underlying CGKA scheme satisfies FS and PCS. We use this fact to conclude that our protocol also satisfies the two properties, while also achieving selective message access. For sender anonymity, the analysis relies on the assumption that leaf secret chosen during key updates of TreeKEM is uniformly random.

We assume that the hash function $\hash$, used in both TreeKEM and our protocol, functions as a pseudorandom generator (PRG). This implies that if $\hash$ receives random input, its output will be uniformly random. Furthermore, we assume the public key encryption (PKE) scheme employed is IND-CPA secure, which guarantees that an adversary cannot distinguish between the ciphertexts of any two chosen plaintexts. The formal security definition for these cryptographic primitives is included in Appendix~\ref{appendix:primitives}. 

Also, a CGKA scheme should already satisfy the following two security properties, according to the security definition of key indistinguishability by Alwen et al.~\cite{alwen2020security}.
To formalize, we first define the epoch $t$, which is a protocol execution counter that advances whenever a control message is processed. Let $\gamma_t$ denotes the state at epoch $t$ and $k_t$ denotes the group key at epoch $t$, we have the following relation for each group member: $(\gamma_t, k_t) \gets \proc(\gamma_{t-1}, T)$ for any legitimate control message $T$.

\begin{itemize}
    \item \textbf{Forward Secrecy}: For an external adversary who has access to all control messages $T$ and compromises a member's state $\gamma_t$, the adversary should not be able to distinguish any key $k_i$ for $i < t$ from a uniform random distribution.
    \item \textbf{Post-Compromise Security}: For an external adversary who has access to all control messages $T$ and compromises a member's state, but a group member successfully creates a commit at epoch $t$ without the adversary's control, the adversary should not be able to distinguish any key $k_i$ for $i > t$ from a uniform random distribution.
\end{itemize}

\myparagraph{Forward Secrecy.}
Assuming that no chatbot is compromised, we show that for an external adversary with access to messages $T$ and a compromised member state $\gamma_t$, any key $k_i$ for $i < t$ remains indistinguishable from uniform random. First, the adversary gains no meaningful information about previous keys from the compromised keys because the underlying CGKA protocol ensures forward secrecy, making any key $k$ generated before the compromise indistinguishable from random. By the definition of PRG, the group keys $k’$ generated before the compromise also retain this indistinguishability, as $k’ = \prg(k)$. Second, the adversary gains no meaningful information from control messages containing encrypted previous keys, as the IND-CPA security of the PKE scheme ensures that these ciphertexts reveal no meaningful information about the plaintext keys and are indistinguishable from encryptions of random values.

\myparagraph{Post-Compromise Security.}
Assuming that no chatbot is compromised, we show that for an external adversary with access to messages $T$ and the state of a compromised member, if a group member performs an uncompromised operation at epoch $t$, then the subsequent key $k_i$ for $i > t$ is indistinguishable from a uniform random distribution. This claim is based on principles similar to those supporting FS, where post-compromise group keys generated by a PRG are indistinguishable from random. In addition, the IND-CPA security of the PKE scheme ensures that control messages reveal no meaningful information about the future keys to an attacker.

\myparagraph{Sender Anonymity.}
Assuming that the new leaf secret chosen during key updates of TreeKEM is uniformly random, we show that a chatbot adversary cannot significantly distinguish any message bundle between any two group members. Information in the message bundle comprises a constant chatbot identifier, a group public key derived from a uniformly random leaf secret via PRG operations, and a ciphertext secured under IND-CPA is indistinguishable from random. 

\myparagraph{Selective Message Access.}
We show that a chatbot adversary within a group cannot distinguish any unauthorized key $k$ from a uniform random distribution by adhering to both forward secrecy and post-compromise security. For a chatbot adversary $\adv$ authorized with key $k_t$ at epoch $t$, forward secrecy ensures that all previous keys $k_i$ for $i<t$ are secure and indistinguishable from random because they are generated via a PRG from a uniformly random source. In addition, these keys maintain indistinguishability under IND-CPA security because the control messages do not contain ciphertexts encrypted with the public key of $\adv$. On the other hand, post-compromise security protects all subsequent keys $k_i$ for $i>t$, preventing $\adv$ from obtaining knowledge of these unauthorized keys and thus protecting future communications within the group.

\subsection{Performance Analysis}

\begin{table*}[htbp]
  \centering
    \begin{small}
    \begin{tabular}{cc|ccc | ccc }
    \hline
    & & \multicolumn{3}{c|}{number of exponentiations} & \multicolumn{3}{c}{number of encryptions or hashes} \\
    & & sender & per user & per chatbot & sender & per user & per chatbot\\
    \hline
    \multirow{2}{*}{Sender Keys}
        & setup & $\bigO{n+m}$ & $\bigO{n+m}$ & $\bigO{n+m}$ & $\bigO{n+m}$ & $\bigO{n+m}$ & $\bigO{n+m}$ \\
        & ongoing & $0$ & $0$ & $0$ & $\bigO{1}$ & $\bigO{1}$ & $\bigO{1}$ \\
    \hline
    \multirow{2}{*}{(A)ART}
        & setup   & $\bigO{n+m}$ & $\bigO{\log(n+m)}$ & $\bigO{\log(n+m)}$ & 0 & 0 & 0 \\
        & ongoing & $\bigO{n+m}$ & $\bigO{n+m}$ & $\bigO{n+m}$ & $\bigO{1}$ & $\bigO{1}$ & $\bigO{1}$ \\
    \hline
    \multirow{2}{*}{MLS}
        & setup   & $\bigO{n+m}$ & $\bigO{1}$ & $\bigO{1}$ & $\bigO{n+m}$ & $\bigO{\log(n+m)}$ & $\bigO{\log(n+m)}$ \\
        & ongoing & $\bigO{\log(n+m)}$ & $\bigO{1}$ & $\bigO{1}$ & $\bigO{\log(n+m)}$ & $\bigO{\log(n+m)}$ & $\bigO{\log(n+m)}$ \\
    \hline
    \multirow{2}{*}{Ours}
        & setup*   & $\bigO{n+m}$ & $\bigO{1}$ & $\bigO{1}$ & $\bigO{n+m}$ & $\bigO{\log n}$ & $\bigO{1}$ \\
        & ongoing & $\bigO{\log n + m}$ & $\bigO{1}$ & $\bigO{1}$ & $\bigO{\log n + m}$ & $\bigO{\log n}$ & $\bigO{1}$ \\
    \hline
    \end{tabular}
    \vspace{1ex}
    
    {\centering *: The overhead for registering pseudonyms, which is equivalent to sending a message from each user to the chatbot, is omitted. \par}
    \end{small}
    \caption{Computation complexity comparison. $n$ = number of group members, $m$ = number of chatbots.}
    \label{table:simple_analysis}
\end{table*}

This section analyzes the overhead of our protocol, both theoretically and empirically, and compares it to other secure group messaging protocols.

\subsubsection{Theoretical Performance Analysis}
\label{sec:eva:the}

In our protocol, the setup phase for a group with $n$ users and $m$ chatbots involves $\bigO{n+m}$ public-key encryption (PKE) and hash operations, where the TreeKEM of user subtree construction requires $\bigO{n}$ PKE and hashes, and computing shared keys for each chatbot takes $\bigO{m}$ PKE and hashes.
Sending a message to chatbots requires $\bigO{\log n + m}$ complexity, where updating user subtree requires $\bigO{\log n}$ PKE and hashes, and computing shared keys for each chatbot takes $\bigO{m}$ PKE and hashes. Pseudonymity adds minimal overhead. The storage requirement is $\bigO{n+m}$ for users and $\bigO{1}$ for chatbots. 
The comparative result is presented in Table~\ref{table:simple_analysis}, and a more detailed analysis is presented in Appendix~\ref{appendix:full_analysis}.

\subsubsection{Experimental Performance Evaluation}
\label{sec:eva:impl}

We implemented \cmrt using Go. Our prototype implementation 
includes two versions: one extends the existing \texttt{libsignal} library~\cite{libsignal}, where individual chats follow the Signal Protocol and group chats follow the Sender Keys Protocol; the other version is based on the Messaging Layer Security (MLS) protocol, using the \texttt{go-mls} library~\cite{go-mls}.
We conducted our experiments on a Mac mini equipped with Apple M2 processor and 16GB of RAM.

Figure~\ref{fig:add-line-plot} shows the time required to add a chatbot to groups of different sizes, measured from the initiation of the addition process to its completion for all users. Regardless of the underlying protocol, adding a chatbot to a group of 50 members takes about 2 milliseconds in our protocol, while it takes about 30 milliseconds in both original protocols. 
Adding a pseudonymous chatbot to a group incurs higher overhead due to pseudonym registration, but this can be minimized by combining registration with the user's first message to the chatbot, avoiding extra roundtrips.

Figure~\ref{fig:send-line-plot} shows the time required to send a message to group members and chatbots, from the start of encryption to the point where all users and chatbots have decrypted the message. 
Sending a message to 50 members and 10 chatbots takes about 10 milliseconds when integrated with MLS and about 5 milliseconds when integrated with the Sender Keys Protocol. 
The overhead of sending a message increases linearly with the number of chatbots, which is consistent with the theoretical analysis. Pseudonymity introduces a small additional overhead due to the signature processes involved. 

To evaluate the performance of our protocol on resource-constrained devices, we also benchmarked chatbot addition and message encryption on low-end containers, as shown in Figure~\ref{fig:add-line-plot-cpu} and Figure~\ref{fig:send-line-plot-cpu} in the appendix. We conclude that our messaging protocol does not introduce prohibitive overhead for users or chatbots, even on lower-end devices. A more detailed analysis and the full results of our experiment are presented in Appendix~\ref{appendix:full_experiment}.

\begin{figure}[htbp]
\captionsetup[subfigure]{aboveskip=-0.5pt}
    \centering
        \begin{subfigure}[b]{0.45\columnwidth}
        \resizebox{\columnwidth}{!}{
            \begin{tikzpicture}
            \scalefont{0.4}
            \begin{axis}[
            sharp plot,
            xmode=normal,
            xlabel=Number of Members,
            ylabel=Time (ms),
            width=6cm, height=4cm,
            xmin=0,xmax=50,
            ymin=0, ymax=220,
            xtick={5,10,15,20,25,30,35,40,45,50},
            ytick={50, 100, 150, 200},
            xlabel near ticks,
            ylabel near ticks,
            ymajorgrids=true,
            grid style=dashed,
            legend style={at={(0.25, 1)},anchor=north},
            ]
    
            \addplot[mark=*, smooth, color=color1] coordinates {
                (5,3.75)
                (10,6.31)
                (15,8.79)
                (20,11.1)
                (25,13.7)
                (30,16.3)
                (35,18.7)
                (40,22.5)
                (45,26.9)
                (50,30.3)
            };
            \addlegendentry{Original}
            
            \addplot[mark=triangle*, smooth, color=color2] coordinates { 
                (5,0.345)
                (10,0.449)
                (15,0.57)
                (20,0.628)
                (25,0.717)
                (30,0.874)
                (35,1.1)
                (40,1.06)
                (45,1.24)
                (50,1.76)
            };
            \addlegendentry{IGA}
            
            \addplot[mark=o, smooth, color=color3] coordinates {
                (5,4.85)
                (10,13.0)
                (15,22.2)
                (20,37.1)
                (25,51.7)
                (30,69.1)
                (35,101.0)
                (40,128.0)
                (45,177.0)
                (50,214.0)
            };
            \addlegendentry{Pseudo.}
    
            \end{axis}
            \end{tikzpicture}
        }
        \caption{Sender Keys Protocol}
    \end{subfigure}
    \begin{subfigure}[b]{0.45\columnwidth}
        \centering
        \resizebox{\columnwidth}{!}{
            \begin{tikzpicture}
            \scalefont{0.4}
            \begin{axis}[
            sharp plot,
            xmode=normal,
            xlabel=Number of Members,
            ylabel=Time (ms),
            width=6cm, height=4cm,
            xmin=0,xmax=50,
            ymin=0, ymax=220,
            xtick={5,10,15,20,25,30,35,40,45,50},
            ytick={50, 100, 150, 200},
            xlabel near ticks,
            ylabel near ticks,
            ymajorgrids=true,
            grid style=dashed,
            ]
    
            \addplot[mark=*, smooth, color=color1] coordinates {
                (5,3.8)
                (10,4.2)
                (15,4.41)
                (20,5.2)
                (25,4.85)
                (30,5.24)
                (35,6.55)
                (40,10.5)
                (45,14.3)
                (50,21.9)
            };
            
            \addplot[mark=triangle*, smooth, color=color2] coordinates { 
                (5,0.647)
                (10,0.823)
                (15,0.874)
                (20,1.03)
                (25,1.16)
                (30,1.31)
                (35,1.53)
                (40,1.82)
                (45,2.27)
                (50,2.93)
            };
            
            \addplot[mark=o, smooth, color=color3] coordinates {
                (5,88.2)
                (10,201.0)
                (15,326.0)
            };
    
            \end{axis}
            \end{tikzpicture}
        }
        \caption{MLS}
    \end{subfigure}
    \vspace{-0.8\baselineskip}
    \caption{Adding Chatbot with Sender Anonymity}
    \label{fig:add-line-plot}
\end{figure}

\begin{figure}[htbp]
    \captionsetup[subfigure]{aboveskip=-1pt}
    \centering
    \begin{subfigure}[b]{0.45\columnwidth}
        \resizebox{\columnwidth}{!}{
            \begin{tikzpicture}
            \scalefont{0.55}
            \begin{axis}[
            sharp plot,
            xmode=normal,
            xlabel=Numbers of Chatbots,
            ylabel=Time (ms),
            width=6cm, height=4cm,
            xmin=0,xmax=50,
            ymin=0, ymax=50,
            xtick={1,5,10,15,20,25,30,35,40,45,50},
            ytick={10, 20, 30, 40, 50},
            xlabel near ticks,
            ylabel near ticks,
            ymajorgrids=true,
            grid style=dashed,
            legend style={at={(0.25, 1)},anchor=north},
            ]
    
            \addplot[
                mark=*,
                smooth,
                color=color1
            ] coordinates {
                (1,0.932)
                (5,1.08)
                (10,1.33)
                (15,1.63)
                (20,1.98)
                (25,2.35)
                (30,2.76)
                (35,3.26)
                (40,3.76)
                (45,4.31)
                (50,4.89)
            };
            \addlegendentry{Original}
            \addplot[
                mark=triangle*,
                smooth,
                color=color2
            ] coordinates {
                (1,3.28)
                (5,4.0)
                (10,5.3)
                (15,7.07)
                (20,9.25)
                (25,11.9)
                (30,15.1)
                (35,18.8)
                (40,22.9)
                (45,27.5)
                (50,32.7)
            };
            \addlegendentry{IGA}
            \addplot[
                mark=o,
                smooth,
                color=color3
            ] coordinates {
                (1,3.32)
                (5,4.21)
                (10,5.72)
                (15,7.67)
                (20,10.0)
                (25,12.9)
                (30,16.3)
                (35,20.2)
                (40,24.6)
                (45,29.4)
                (50,34.7)
            };
            \addlegendentry{Pseudo.}
    
            \end{axis}
            \end{tikzpicture}
        }
        \caption{Sender Keys Protocol}
     \end{subfigure}
    \begin{subfigure}[b]{0.45\columnwidth}
        \resizebox{\columnwidth}{!}{
            \begin{tikzpicture}
            \scalefont{0.55}
            \begin{axis}[
            sharp plot,
            xmode=normal,
            xlabel=Numbers of Chatbots,
            ylabel=Time (ms),
            width=6cm, height=4cm,
            xmin=0,xmax=50,
            ymin=0, ymax=50,
            xtick={1,5,10,15,20,25,30,35,40,45,50},
            ytick={10, 20, 30, 40, 50},
            xlabel near ticks,
            ylabel near ticks,
            ymajorgrids=true,
            grid style=dashed,
            legend style={at={(0.24, 1)},anchor=north},
            ]

            \addplot[
                mark=*,
                smooth,
                color=color1
            ] coordinates {
                (1,8.7)
                (5,9.96)
                (10,11.4)
                (15,13.3)
                (20,16.3)
                (25,19.5)
                (30,23.4)
                (35,27.5)
                (40,31.9)
                (45,36.9)
                (50,42.3)
            };
            \addplot[
                mark=triangle*,
                smooth,
                color=color2
            ] coordinates {
                (1,8.69)
                (5,9.67)
                (10,11.3)
                (15,13.5)
                (20,16.2)
                (25,19.2)
                (30,22.8)
                (35,26.8)
                (40,31.4)
                (45,36.3)
                (50,41.6)
            };
            \addplot[
                mark=o,
                smooth,
                color=color3
            ] coordinates {
                (1,8.72)
                (5,9.85)
                (10,11.7)
                (15,14.1)
                (20,16.9)
                (25,20.1)
                (30,23.9)
                (35,28.1)
                (40,32.8)
                (45,38.0)
                (50,43.7)
            };

            \end{axis}
            \end{tikzpicture}
        }
        \caption{MLS}
    \end{subfigure}
    \vspace{-0.8\baselineskip}
    \caption{Sending Messages with Sender Anonymity}
    \label{fig:send-line-plot}
\end{figure}

%% file: sections/discussions.tex
\section{Discussions}

\subsection{Integrating Other Privacy Mechanisms}
Our modular design supports seamless integration with existing privacy solutions for group messaging. 
The user subtree component of \cmrtt is built on an abstraction of the CGKA scheme, allowing it to be replaced by other CGKA-based protocols, provided they satisfy the requirements for FS and PCS.
For instance, integrating administrated CGKA~\cite{balbas2023cryptographic} could additionally strengthen group administrative controls.
Metadata-hiding techniques can also be employed to enhance user privacy against group outsiders. For instance, applying CGKA-based membership-hiding methods~\cite{hashimoto2022hide, emura2022membership}, which encrypts and authenticates group metadata using the shared group key, can ensure membership anonymity. Similarly, applying MLS’s Private Message~\cite{rfc9420}, which encrypts control messages, can conceal sender identities.

\subsection{Limitations of \cmrt}
The effectiveness of \cmrt heavily relies on the accuracy of the trigger function in determining the relevance of messages. A poorly designed or maliciously crafted trigger function could mark all messages as relevant, granting a chatbot unrestricted access despite \cmrt's protections. Addressing this issue may be beyond the scope of the messaging protocol itself. Instead, a vetting process, either manual or automated, could be implemented to mitigate such abuses of trigger functions.
Additionally, while we identify two key privacy properties against overprivileged chatbots, namely selective message access and sender anonymity, the list may not be exhaustive. As new risks emerge and additional privacy properties are proposed, \cmrt can serve as a foundation for future refinements.

\subsection{Roadmap Toward a Privacy-Preserving Solution for Chatbots}
Developing privacy-preserving chatbots for group messaging platforms is a complex challenge that involves both technical and user-centric considerations. While \cmrt provides a foundational framework by addressing two key privacy properties at the protocol level, there remain broader areas for exploration. %

\myparagraph{Enhancing Usability.} One important direction is to improve the usability of privacy-preserving chatbots. This includes designing intuitive interfaces that inform users of the chatbot's presence, clearly communicate chatbot permissions, and enable seamless workflows for granting or restricting access. For example, Keybase displays chatbot permissions on installation pages and notifies users that a chatbot can read their messages by displaying an icon next to those messages~\cite{keybasebasic}. However, to our knowledge, no user studies have been conducted to evaluate the effectiveness of design approaches for communicating chatbot permissions.

\myparagraph{Designing Permission Model.} 
Refining permission models for chatbots is another important area of research to enhance privacy protections.
Previous research on Android permissions has explored automated, context-aware permission decision support~\cite{wijesekera2018contextualizing, gao2019autoper}. Building on this approach, future work could focus on developing dynamic and adaptive permission frameworks designed for evolving group dynamics and user needs. These frameworks could integrate natural language processing (NLP) techniques to help users decide whether a chatbot can read certain messages or a message should be sent anonymously, thereby reducing users' cognitive load.

\myparagraph{Understanding User Perceptions.} Understanding users perceptions, acceptance, and trust of privacy-preserving chatbots is critical. 
Several studies have explored users' privacy concerns and perceptions of chatbots in one-on-one interactions~\cite{gumusel2024literature, ischen2020privacy, gieselmann2023more}, but group chat settings remain underexplored. 
Understanding how users perceive privacy features, what concerns they prioritize, and how they interact with these systems is essential to improving adoption and engagement.

%% file: sections/related_work.tex
\section{Related Work}
This work presents a secure group messaging protocol protecting users' privacy from chatbots. In the previous sections, we reviewed secure messaging protocols and messaging platforms supporting chatbots. This section further considers related work on chatbots' security and privacy issues and permission frameworks in smart devices.

\subsection{Chatbot Security}
\label{ssec:related:chatbot}

Several studies have conducted large-scale security evaluations of chatbots on modern messaging platforms. Edu et al.~\cite{edu2022exploring} analyzed over 15,000 Discord chatbots 
and found that over 40\% of the chatbots examined ask for permission to access message history, but less than 5\% of them offer a privacy policy.
Similarly, Chen et al.~\cite{chen2022experimental} analyzed design flaws in chatbot-like third-party apps on Business Collaboration Platforms (BCP), such as Slack. %
Their analysis showed that these apps can steal messages or impersonate users. %
The use of runtime policy checks and explicit user confirmation are suggested as countermeasures.
These studies underscore the privacy risks associated with chatbots, providing strong motivation for our work.

Biswas~\cite{biswas2020privacy} proposed methods for service providers to filter encrypted messages sent to chatbots using searchable encryption~\cite{boneh2004public}. While their objective aligns with our goal of selective message access, their approach lacks robust E2EE properties like FS. Nonetheless, their work highlights the potential for achieving selective message access through service providers while protecting message confidentiality.

\subsection{Permission Frameworks in Smart Devices} 
\label{ssec:related:permission}
Mobile and web app permissions are well-studied, with findings showing that user-consent models often fail to inform users effectively about privacy risks. Chia et al.~\cite{chia2012app} conducted a large-scale study on Facebook apps, Chrome extensions, and Android apps. They revealed that many apps use misleading tactics to request excessive permissions. Similarly, Felt et al.~\cite{felt2011android, felt2012android} found Android's permission system ineffective because developers often fail to follow the least privilege principle, leading to over-privileged apps, and users often ignore or misunderstand permission warnings, leading to uninformed decisions. These challenges highlight the importance of ensuring that the trigger function in our protocol is both developer-friendly and capable of clearly communicating permissions to users. 

Smart speakers, like chatbots that constantly listen to messages, face similar challenges due to their always-on microphones. Manikonda et al.~\cite{manikonda2018s} found that users reported heightened privacy concerns after learning about the always-listening nature of smart speakers. Lau et al.~\cite{lau2018alexa} identified widespread user misunderstandings and limited use of privacy controls. Moreover, both Dubois et al.~\cite{dubois2020speakers} and Schönherr et al.~\cite{schonherr2022exploring} demonstrated that smart speakers, when actively listening for activation words, can be unintentionally triggered, leading to unintended recordings. 
Chatbots in group chats, which may also operate in an always-on mode, could raise similar concerns when chatbots are mistakenly triggered.

%% file: sections/conclusion.tex
\section{Conclusion}
This paper identifies two key privacy issues in group messaging protocols involving chatbots, and highlights the lack of existing platform that adequately addresses these issues while preserving robust E2EE.
To address these challenges, we introduce selective message access and two forms of sender anonymity, and propose a secure group messaging protocol called \cmrt, which efficiently manages multiple keys among group members and chatbots while ensuring sender anonymity.
The theoretical analysis confirms the possibility of achieving the claimed properties without significant overhead, and our implementation demonstrates both its efficiency and its seamless integration into existing secure messaging services.
Finally, we present a roadmap toward a more comprehensive solution to this problem, aiming to raise awareness and inspire future research on this critical issue.

%% file: sections/ethics.tex
\section*{Ethical Consideration}
Our study is based on two publicly available conversation datasets, both of which were collected from public chat rooms and compliant with the policies of the respective messaging services. The datasets are fully anonymized, and we do not attempt to re-identify any member, except a chatbot as clearly discussed in Section~\ref{sec:attack}. When consulting about the chatbot, one of the researchers used their authentic Discord account and asked in a public channel.

Telegram, Slack, and Discord are aware of the chatbot security issues to the extent that they already provide mitigations that do not work with E2EE. For platforms without official chatbot support, such as WhatsApp, disguising chatbots as users is against their terms of service. Therefore, we see no imminent need to inform them about the issues. We are in the process of informing LINE about the issues.

\section*{Open Science}
The analysis scripts for the Pushshift Telegram dataset used in Section~\ref{sec:attack:cross-group}, the basic chatbot implementations for various messaging services discussed in Section~\ref{sec:Case Studies:official}, and the prototype implementation of SnoopGuard mentioned in Section~\ref{sec:eva:impl} are available at \url{https://github.com/csienslab/snoopguard-artifact} and \url{https://zenodo.org/records/14729613}.

%% file: sections/appendix.tex
\appendix

\section{Cryptographic Primitives}\label{appendix:primitives}

\subsection{Pseudorandom Generators}
Let $\mathcal{W}$ be the domain and range of the pseudorandom generator (PRG). A PRG is a function $\prg: \mathcal{W} \rightarrow \mathcal{W}$ that, given an input $U$ uniformly sampled from $\mathcal{W}$, produces an output $\prg(U)$ that is indistinguishable from a uniformly random element $U’ \in \mathcal{W}$. The security of a PRG is measured by the advantage for an attacker $\adv$ has in distinguishing between $\prg(U)$ and $U'$, denoted as $\advantage{\prg}{prg}[(\adv)]$.

\begin{definition}
    A $\prg$ scheme is \textit{secure} if for all efficient adversaries $\adv$ and a security parameter $\secpar$, there is a negligible function $\mathsf{negl}$ such that $\advantage{\prg}{prg}[(\adv)] \le \negl$.
\end{definition}

\subsection{Public Key Encryption}
A public-key encryption (PKE) scheme $\pke=(\pkeg, \pkenc, \pkdec)$ consists of the following algorithms:
\begin{itemize}
    \item $(\sk, \pk) \sample \pkeg(s)$: Generates a PKE key pair from a secret $s$.
    \item $c \sample \pkenc(\pk, m)$: Encrypts the message $m$ using the public key $\pk$ and outputs a ciphertext $c$.
    \item $m \gets \pkdec(\sk, c)$: Decrypts the ciphertext $c$ using $\sk$ and outputs the message $m$.
\end{itemize}

\textit{IND-CPA security.} Consider the following game:
\procedureblock[]{$\indcpa_\pke^\adv(\secpar)$}{ 
    b \sample \bin \\
    (\sk, \pk) \sample \pkeg(\secparam) \\
    (m_0,m_1) \sample \adv(\secparam , \pk) \\
    c \sample \enc(\pk,m_b) \\
    b’ \sample \adv(\secparam, c) \\ 
    \pcreturn 1_{b = b’} 
}

\begin{definition}
    Let \advantage{\pke}{\indcpa}[(\adv)] denote the advantage of adversary $\adv$ winning the $\indcpa$ game, A $\pke$ scheme is $CPA-secure$ if for all efficient adversaries $\adv$ and a security parameter $\secpar$, there is a negligible function $\mathsf{negl}$ such that
    \begin{align*}
        \advantage{\pke}{\indcpa}[(\adv)] \le \negl.
    \end{align*}
\end{definition}

\section{Pseudoprocedure of \cmrt}\label{appendix:pseudo}

This section presents the pseudoprocedure of our group messaging protocol in Figure~\ref{fig:cmrt_protocol}. A user with state $\gamma$ sends a message $m$ by invoking $\usrsend(\gamma, m)$, which generates a control message $T$. Other users update their states by calling $\proc(\gamma, T)$, while only the triggered chatbots can successfully decrypts the message $m$ by invoking $\cbtrecv(\gamma, T)$, where $\gamma$ represents their respective states. Similarly, the chatbots send a message with $\cbtsend$, which generates control message $T$ to be processed by $\usrrecv$. The operations $\setpk$ and $\getpk$ represent incorruptible operations for registering and retrieving metadata associated with a chatbot identified by $\cid$ from the service provider.

\begin{figure*}[htb]
    \begin{footnotesize}
    \begin{pchstack}[center, space=2em]
    \begin{pcvstack}[space=0.5em]
        \procedure[codesize=\scriptsize]{$\init(\id)$} {
            \pcln \gamma.s0 \gets \cgka.\init(\id) \\
            \pcln \gamma.\chatbots[\cdot] \gets \perp \\
            \pcln \pcreturn \gamma
        }
        \procedure[codesize=\scriptsize]{$\creategroup(\gamma, \id_1, \dots, \id_n)$} {
            \pcln (\gamma.s0, T) \gets \cgka.\creategroup(\gamma.s0, \id_1, \dots, \id_n) \\
            \pcln \pcreturn (\gamma, T) 
        }
        \procedure[codesize=\scriptsize]{$\addcbt(\gamma, \cid)$} {
            \pcln \gpk \gets \gamma.s0.\gpk \pccomment{user subtree's pk} \\
            \pcln (\pk_\cid, f_\cid, s) \gets \getpk(\cid) \\
            \pcln k \sample \bin^\secpar; (\csk, \cpk) \gets \pkeg(k) \\
            \pcln e \gets \pkenc(\pk_\cid, k) \\
            \pcln T \gets (\addcbt, \cid, \gpk, \cpk, e) \\
            \pcln \pcreturn (\gamma, T)
        }
        \procedure[codesize=\scriptsize]{$\mathsf{proc}(\gamma, T=(\addcbt, \cid, \gpk, \cpk, e))$} {
            \pcln (\pk_\cid, f_\cid, s) \gets \getpk(\cid) \\
            \pcln \pcassert \verify(\pk_\cid, s, f_\cid) \pccomment{checks integrity} \\
            \pcln \gamma.\chatbots[\cid] \gets (f_\cid, \gamma.s0.\gsk, \cpk) \\
            \pcln \pcreturn \gamma
        }
        \procedure[codesize=\scriptsize]{$\remcbt(\gamma, \cid)$} {
            \pcln T \gets (\remcbt, \cid) \\
            \pcln \pcreturn (\gamma, T)
        }
        \procedure[codesize=\scriptsize]{$\mathsf{proc}(\gamma, T=(\remcbt, \cid))$} {
            \pcln \gamma.\chatbots[\cid] \gets \perp \\
            \pcln \pcreturn \gamma
        }
    \end{pcvstack}
    \begin{pcvstack}[space=1em]
        \procedure[codesize=\scriptsize]{$\usrsend(\gamma, m)$} {
            \pcln (\gamma.s0, T_0), \gets \cgka.\upd(\gamma.s0) \\
            \pcln (\gamma.s0, k) \gets \cgka.\proc(\gamma.s0, T_0) \\ 
            \pcln (\gsk, \gpk) \gets \pkeg(k) \\
            \pcln k' \gets \hash(k) \\
            \pcln c \gets \enc(k', m) \pccomment{encrypts the message} \\
            \pcln T \gets (T_0, c, \gpk) \\
            \pcln \pcfor \cid : f_\cid(m) = 1 \\
            \pcln \t \cpk \gets \gamma.\chatbots[\cid].\cpk \\
            \pcln \t e \gets \pkenc(\cpk, k') \\
            \pcln \t T \gets T \  \Vert \  (\cid, e) \\
            \pcln \t \gamma.\chatbots[\cid].\gsk \gets \gsk \\
            \pcln \pcreturn (\gamma, T)
        }
        \procedure[codesize=\scriptsize]{$\mathsf{proc}(\gamma, T=(T_0, c, \gpk, (\cid_i, e_i)_{i=1\dots n}))$} {
            \pccomment{This processes $T$ generated by $\usrsend$} \\
            \pcln (\gamma.s0, k) \gets \cgka.\proc(\gamma.s0, T_0) \\
            \pcln (\gsk, \gpk) \gets \pkeg(k) \\
            \pcln \pcfor i = 1\dots n \\
            \pcln \t \gamma.\chatbots[\cid_i].\gsk \gets \gsk \\
            \pcln \pcreturn \gamma
        }
        \procedure[codesize=\scriptsize]{$\usrrecv(\gamma, T=(\cid, c, e, \cpk))$} {
            \pcln k' \gets \pkdec(\gamma.\chatbots[\cid].\gsk, e) \\
            \pcln m \gets \dec(k', c) \\
            \pcln \gamma.\chatbots[\cid].\cpk \gets \cpk \\
            \pcln \pcreturn (\gamma, m)
        }
    \end{pcvstack}
    \begin{pcvstack}[space=0.5em, boxed]
        \procedure[codesize=\scriptsize]{$\init(\cid, f_\cid, (\sk_\cid, \pk_\cid))$} {
            \pcln \gamma.\me \gets \cid \\
            \pcln \gamma.\gpk \gets \perp \\
            \pcln (\gamma.\sk_\cid, \gamma.\pk_\cid) \gets (\sk_\cid, \pk_\cid) \\
            \pcln (\gamma.\csk, \gamma.\cpk) \gets (\perp, \perp) \\
            \pcln s \gets \sign(\sk_\cid, f_\cid) \\
            \pcln \setpk(\cid, \pk_\cid, f_\cid, s) \\
            \pcln \pcreturn \gamma
        }
        \procedure[codesize=\scriptsize]{$\proc(\gamma, T=(\addcbt, \cid, \gpk, \cpk, e))$} {
            \pcln \gamma.\gpk \gets \gpk \\
            \pcln k \gets \pkdec(\gamma.\sk_\cid, e) \\
            \pcln (\gamma.\csk, \gamma.\cpk) \gets \pkeg(k) \\
            \pcln \pcreturn \gamma
        }
        \procedure[codesize=\scriptsize]{$\cbtsend(\gamma, m)$} {
            \pcln k \sample \bin^\secpar \\
            \pcln (\gamma.\csk, \gamma.\cpk) \gets \pkeg(k) \\
            \pcln k' \gets \hash(k) \\
            \pcln c \gets \enc(k', m) \pccomment{encrypts the message} \\
            \pcln e \gets \pkenc(\gamma.\gpk, k') \\
            \pcln T \gets (\gamma.\me, c, e, \gamma.\cpk) \\
            \pcln \pcreturn (\gamma, T)
        }
        \procedure[codesize=\scriptsize]{$\cbtrecv(\gamma, T=(T_0, c, \gpk, (\cid_i, e_i)))$} {
            \pccomment{Only processes $(\cid_i, e_i)$ where $\cid_i=\gamma.\me$} \\
            \pcln k' \gets \pkdec(\gamma.\csk, e_i) \\
            \pcln m \gets \dec(k', c) \\
            \pcln \gamma.\gpk \gets \gpk \\
            \pcln \pcreturn (\gamma, m)
        }
    \end{pcvstack}
    \end{pchstack}
    \end{footnotesize}
    \caption{The \cmrt protocol. Unboxed algorithms are used by users while \framebox{boxed} algorithms are used by chatbots.}
    \label{fig:cmrt_protocol}
\end{figure*}

\section{Theoretical Performance Analysis of the Secure Group Messaging}
\label{appendix:full_analysis}
This section extends Section~\ref{sec:eva:the} by presenting a more detailed analysis of the theoretical performance of our secure group messaging protocol.

\myparagraph{Baseline.} For setup phase, the group initiator uses $\bigO{n+m}$ PKE operations and $\bigO{n+m}$ symmetric operations. The construction of the TreeKEM with $n$ members involves $\bigO{n}$ PKE operations and hash operations, respectively. To compute the shared secret for each chatbot, the initiator also performs $\bigO{m}$ PKE operations and hash operations, respectively. 
For the receivers, each user requires $\bigO{1}$ PKE operations and $\bigO{\log n}$ hash operations to initiate the TreeKEM. 
For each chatbot, it takes $\bigO{1}$ PKE operations to decrypt the secret, but only $\bigO{1}$ hash operations to compute the shared secret due to the unbalanced tree structure. 

Suppose a user sends a message to the chatbots. The message sender performs $\bigO{\log n + m}$ PKE operations and symmetric operations, respectively. Updating the TreeKEM involves $\bigO{\log n}$ public key generations and hash operations.
Each chatbot takes $\bigO{1}$ PKE operations and hash operations for the sender to do the key update and message encryption, respectively, and there are $m$ chatbots, imposing $\bigO{m}$ overhead. 
Message recipients, including users and chatbots, perform identical actions as in the setup phase to update the secret, resulting in the same overhead. 

Adding a chatbot to the group is almost the same as sending a message to a chatbot. The initiator performs $\bigO{\log n}$ PKE operations and symmetric operations, respectively, to update the TreeKEM. Both the chatbot and other members perform the same actions as in the setup phase to update the secret.

\myparagraph{Pseudonymity.} For the setup phase, registering a pseudonym is equivalent to sending a message to the chatbot. For the ongoing phase, using pseudonyms requires both sender and receiver $\bigO{1}$ additional PKE operations to create and verify the signature. This does not affect the overall complexity. 

\myparagraph{Trigger Concealment.} The sender fakes key updates using random bytes and sends them to chatbots that should not receive the message. This is equivalent to triggering all chatbots, and therefore maintaining the same complexity for the ongoing phase. 

\myparagraph{Storage Overhead.} The storage overhead for each user is $\bigO{n+m}$, which includes $\bigO{n}$ keys for the TreeKEM and $\bigO{m}$ keys for all the chatbots. This is equivalent to the MLS with $n+m$ members. However, each chatbot only needs to store $\bigO{1}$ public key for users' subtree. This represents an advantage compared to MLS, which requires $\bigO{m+n}$ storage for each chatbot.

\section{Experimental Performance Evaluation of the Secure Group Messaging}
\label{appendix:full_experiment}

This section extends Section~\ref{sec:eva:impl} and presents the full results of our experiments, as well as a more detailed analysis of the results.

\myparagraph{Adding a Chatbot.}
To demonstrate that adding a chatbot has an acceptable overhead, we measure the time from the initiation of the invitation to the chatbot until all members and chatbots complete the necessary key exchanges. In the case of a pseudonymous chatbot, all users register their pseudonyms. 

Figure~\ref{fig:add-line-plot} illustrates the time spent adding a chatbot to a group with varying group sizes, based on different levels of sender anonymity. For comparison, we also include the traditional scenario where the chatbot is treated as a user. For the Signal Protocol (Sender Keys Protocol), adding a chatbot to an IGA-secure group takes significantly less time because there is no need to distribute sender keys. For the MLS protocol, the original protocol takes slightly more time due to the larger group size, as chatbots are counted as members and therefore cause more overhead when adding members. %

\myparagraph{Sending a Message.}
The timer starts when the sender's client takes the plaintext and stops when the last receiver decrypts and outputs the message content. For each experiment, we maintain a fixed group size of 50 members and measure the time it takes for all group members and chatbots to receive the message. We assume that all chatbots will receive the message regardless of the content, showing the worst-case scenario. %
Our experiments focus on the efficiency of our protocols with respect to the number of chatbots. We did not consider the effect of group size because we use the underlying protocol for group members, and the efficiency of the Sender Keys Protocol and MLS are not relevant in our experiments.  

Figure~\ref{fig:send-line-plot} shows the time required to send a message to group members and chatbots with varying levels of sender anonymity, as well as whether to conceal triggers from the service provider. We also include the traditional scenario for comparison. 
Consistent with the theoretical analysis, our protocol introduces overhead linear to the number of chatbots. Pseudonymity results in slightly higher overhead, possibly due to the additional signature processes.

\paragraph{Performance on Resource-Constrained Devices.} 

Most users access messaging services on mobile devices. To demonstrate that our protocol performs well under mobile-like constraints, we simulate its behavior in environments with limited CPU resources. Specifically, we run our protocol in Docker containers with CPU allocations ranging from 0.5 to 1.0 CPUs. Each container was benchmarked using Geekbench 6~\cite{geekbench}, a widely-used tool for assessing mobile device performance, to understand how the allocated resources perform. 
After establishing the performance baseline, we simulate the protocol in the constrained environment. For the chatbot addition experiment, we measure the time required for the device to generate all the necessary information, complete the key exchange, and register a pseudonym, if applicable. 
For the message sending scenario, we compute the time required for the device to generate all ciphertexts of the messages and, if necessary, the required signatures.
Unlike previous experiments that evaluate overall system performance, this setup isolates the computation performed on a single device to evaluate the protocol's suitability for resource-constrained environments.

The CPU constraints used in our experiments resulted in Geekbench 6 single-core scores ranging from 1,171 to 2,629. These scores represent a spectrum from budget smartphones (e.g., Qualcomm Snapdragon 865, MediaTek Dimensity 8100) to high-end devices (e.g., Apple M2, Apple A16) as of 2024.

Figure~\ref{fig:add-line-plot-cpu} shows the results of the chatbot addition experiment for a group with 50 members and 30 chatbots. Without pseudonymity, the overhead remains below 15 ms for all CPU configurations. With pseudonymity enabled, the overhead increases slightly due to pseudonym generation, but remains below 20 ms even on the most constrained CPU setting. Figure~\ref{fig:send-line-plot-cpu} shows the results of the message sending experiment for the same group configuration. Using the Signal protocol, the time to generate ciphertexts remains below 5 ms, even with the lowest CPU allocation. While MLS introduces more overhead, it remains below 10 ms for all configurations.  

These results show that \cmrt performs efficiently across devices, from low-end to high-end, with minimal latency even under significant CPU constraints, making it suitable for diverse mobile devices.

\begin{figure}[htbp]
\captionsetup[subfigure]{aboveskip=-0.5pt}
    \centering
        \begin{subfigure}[b]{0.45\columnwidth}
        \resizebox{\columnwidth}{!}{
            \begin{tikzpicture}
            \scalefont{0.4}
            \begin{axis}[
            sharp plot,
            xmode=normal,
            xlabel={Geekbench Single-Core Score},
            ylabel={Time (ms)},
            width=6cm, height=4.5cm,
            xmin=1000,xmax=3000,
            ymin=0, ymax=25,
            ytick={5, 10, 15, 20, 25},
            xlabel near ticks,
            ylabel near ticks,
            ymajorgrids=true,
            grid style=dashed,
            legend style={at={(0.25, 1)},anchor=north},
            ]
    
            \addplot[mark=*, smooth, color=color1] coordinates {(1171,12.6) (1485,10.9) (1805,10.7) (2104,10.1) (2409,10.4) (2629,10.5) };
            \addlegendentry{Original}
            
            \addplot[mark=triangle*, smooth, color=color2] coordinates {(1171,0.643) (1485,0.583) (1805,0.508) (2104,0.409) (2409,0.387) (2629,0.364) };
            \addlegendentry{IGA}
            
            \addplot[mark=o, smooth, color=color3] coordinates {(1171,1.9) (1485,1.72) (1805,1.49) (2104,1.44) (2409,1.27) (2629,1.23) };
            \addlegendentry{Pseudo.}
    
            \end{axis}
            \end{tikzpicture}
        }
        \caption{Sender Keys Protocol}
    \end{subfigure}
    \begin{subfigure}[b]{0.45\columnwidth}
        \centering
        \resizebox{\columnwidth}{!}{
            \begin{tikzpicture}
            \scalefont{0.4}
            \begin{axis}[
            sharp plot,
            xmode=normal,
            xlabel={Geekbench Single-Core Score},
            ylabel={Time (ms)},
            width=6cm, height=4.5cm,
            xmin=1000,xmax=3000,
            ymin=0, ymax=25,
            ytick={5, 10, 15, 20, 25},
            xlabel near ticks,
            ylabel near ticks,
            ymajorgrids=true,
            grid style=dashed,
            ]
    
            \addplot[mark=*, smooth, color=color1] coordinates {(1171,2.41) (1485,2.04) (1805,1.71) (2104,1.57) (2409,1.44) (2629,1.4) };
            
            \addplot[mark=triangle*, smooth, color=color2] coordinates {(1171,0.871) (1485,0.65) (1805,0.479) (2104,0.402) (2409,0.479) (2629,0.386) };
            
            \addplot[mark=o, smooth, color=color3] coordinates {(1171,16.2) (1485,14.1) (1805,12.3) (2104,12.9) (2409,11.6) (2629,11.5) };

            \end{axis}
            \end{tikzpicture}
        }
        \caption{MLS}
    \end{subfigure}
    \vspace{-0.8\baselineskip}
    \caption{Adding chatbot with sender anonymity under various CPU constraints}
    \label{fig:add-line-plot-cpu}
\end{figure}

\begin{figure}[htbp]
    \captionsetup[subfigure]{aboveskip=-1pt}
    \centering
    \begin{subfigure}[b]{0.45\columnwidth}
        \resizebox{\columnwidth}{!}{
            \begin{tikzpicture}
            \begin{axis}[
                sharp plot,
                xmode=normal,
                xlabel={Geekbench Single-Core Score},
                ylabel={Time (ms)},
                width=6cm, height=4.5cm,
                xmin=1000,xmax=3000,
                ymin=0, ymax=20,
                xlabel near ticks,
                ylabel near ticks,
                ymajorgrids=true,
                grid style=dashed,
                legend style={at={(0.24, 1)},anchor=north},
                ]
            \addplot[mark=*, smooth, color=color1] plot coordinates {(1171,0.51) (1485,0.437) (1805,0.418) (2104,0.41) (2409,0.405) (2629,0.418) };
            \addlegendentry{Original}
            
            \addplot[mark=triangle*, smooth, color=color2] plot coordinates {(1171,6.64) (1485,6.01) (1805,5.7) (2104,5.55) (2409,5.53) (2629,5.5) };
            \addlegendentry{IGA}
            
            \addplot[mark=o, smooth, color=color3] plot coordinates {(1171,7.71) (1485,7.03) (1805,6.68) (2104,6.63) (2409,6.47) (2629,6.43) };
            \addlegendentry{Pseudo}
            
            \end{axis}
            \end{tikzpicture}
        }
        \caption{Sender Keys Protocol}
     \end{subfigure}
    \begin{subfigure}[b]{0.45\columnwidth}
        \resizebox{\columnwidth}{!}{
            \begin{tikzpicture}
            \begin{axis}[
                sharp plot,
                xmode=normal,
                xlabel={Geekbench Single-Core Score},
                ylabel={Time (ms)},
                width=6cm, height=4.5cm,
                xmin=1000,xmax=3000,
                ymin=0, ymax=20,
                xlabel near ticks,
                ylabel near ticks,
                ymajorgrids=true,
                grid style=dashed,
                legend style={at={(0.24, 1)},anchor=north},
                ]
            \addplot[mark=*, smooth, color=color1] plot coordinates {(1171,14.0) (1485,12.6) (1805,11.8) (2104,11.4) (2409,11.3) (2629,11.3) };
            \addplot[mark=triangle*, smooth, color=color2] plot coordinates {(1171,14.6) (1485,13.0) (1805,12.2) (2104,12.0) (2409,11.9) (2629,11.9) };
            \addplot[mark=o, smooth, color=color3] plot coordinates {(1171,15.7) (1485,14.0) (1805,13.2) (2104,13.0) (2409,12.8) (2629,12.7) };
            
            \end{axis}
            \end{tikzpicture}
        }
        \caption{MLS}
    \end{subfigure}
    \vspace{-0.8\baselineskip}
    \caption{Sending messages with sender anonymity under various CPU constraints}
    \label{fig:send-line-plot-cpu}
\end{figure}

%% file: main.bbl
\begin{thebibliography}{10}

\bibitem{antie_meiyu}
Auntie meiyu, your trusted fact-checking confidant.
\newblock \url{https://checkcheck.me/en/}.
\newblock Accessed on 2024-02-07.

\bibitem{geekbench}
Geekbench 6 - cross-platform benchmark.
\newblock \url{https://www.geekbench.com/}.
\newblock Accessed on 2025-01-02.

\bibitem{alwen2019double}
Jo{\"e}l Alwen, Sandro Coretti, and Yevgeniy Dodis.
\newblock The double ratchet: security notions, proofs, and modularization for the signal protocol.
\newblock In {\em Annual International Conference on the Theory and Applications of Cryptographic Techniques}, pages 129--158. Springer, 2019.

\bibitem{alwen2020security}
Jo{\"e}l Alwen, Sandro Coretti, Yevgeniy Dodis, and Yiannis Tselekounis.
\newblock Security analysis and improvements for the ietf mls standard for group messaging.
\newblock In {\em Annual International Cryptology Conference}, pages 248--277. Springer, 2020.

\bibitem{alwen2021modular}
Jo{\"e}l Alwen, Sandro Coretti, Yevgeniy Dodis, and Yiannis Tselekounis.
\newblock Modular design of secure group messaging protocols and the security of mls.
\newblock In {\em Proceedings of the 2021 ACM SIGSAC Conference on Computer and Communications Security}, pages 1463--1483, 2021.

\bibitem{alwen2020continuous}
Jo{\"e}l Alwen, Sandro Coretti, Daniel Jost, and Marta Mularczyk.
\newblock Continuous group key agreement with active security.
\newblock In {\em Theory of Cryptography}, pages 261--290. Springer, 2020.

\bibitem{balbas2022analysis}
David Balb{\'a}s, Daniel Collins, and Phillip Gajland.
\newblock Analysis and improvements of the sender keys protocol for group messaging.
\newblock In {\em XVII Reuni{\'o}n espa{\~n}ola sobre criptolog{\'\i}a y seguridad de la informaci{\'o}n. RECSI 2022}, volume 265, page~25. Ed. Universidad de Cantabria, 2022.

\bibitem{balbas2023cryptographic}
David Balb{\'a}s, Daniel Collins, and Serge Vaudenay.
\newblock Cryptographic administration for secure group messaging.
\newblock In {\em 32nd USENIX Security Symposium (USENIX Security 23)}, pages 1253--1270, 2023.

\bibitem{rfc9420}
Richard Barnes, Benjamin Beurdouche, Raphael Robert, Jon Millican, Emad Omara, and Katriel Cohn-Gordon.
\newblock {The Messaging Layer Security (MLS) Protocol}.
\newblock RFC 9420, July 2023.

\bibitem{baumgartner2020pushshift}
Jason Baumgartner, Savvas Zannettou, Megan Squire, and Jeremy Blackburn.
\newblock The pushshift telegram dataset.
\newblock In {\em Proceedings of the international AAAI conference on web and social media}, volume~14, pages 840--847, 2020.

\bibitem{chatbot_popularity}
Jeff Beckman.
\newblock 120+ chatbot statistics for 2024 (already mainstream).
\newblock \url{https://techreport.com/statistics/software-web/chatbot-statistics/}.
\newblock Accessed on 2024-09-01.

\bibitem{bhargavan2018treekem}
Karthikeyan Bhargavan, Richard Barnes, and Eric Rescorla.
\newblock {TreeKEM: Asynchronous Decentralized Key Management for Large Dynamic Groups A protocol proposal for Messaging Layer Security (MLS)}.
\newblock Research report, {Inria Paris}, May 2018.

\bibitem{bienstock2022more}
Alexander Bienstock, Jaiden Fairoze, Sanjam Garg, Pratyay Mukherjee, and Srinivasan Raghuraman.
\newblock A more complete analysis of the signal double ratchet algorithm.
\newblock In {\em Annual International Cryptology Conference}, pages 784--813. Springer, 2022.

\bibitem{biswas2020privacy}
Debmalya Biswas.
\newblock Privacy preserving chatbot conversations.
\newblock In {\em IEEE International Conference on Artificial Intelligence and Knowledge Engineering (AIKE)}, 2020.

\bibitem{boneh2004public}
Dan Boneh, Giovanni Di~Crescenzo, Rafail Ostrovsky, and Giuseppe Persiano.
\newblock Public key encryption with keyword search.
\newblock In {\em Advances in Cryptology - EUROCRYPT 2004}, pages 506--522. Springer, 2004.

\bibitem{borisov2004off}
Nikita Borisov, Ian Goldberg, and Eric Brewer.
\newblock Off-the-record communication, or, why not to use pgp.
\newblock In {\em Proceedings of the 2004 ACM workshop on Privacy in the electronic society}, pages 77--84, 2004.

\bibitem{groupbutler}
BotoStore.
\newblock Group butler.
\newblock \url{https://botostore.com/c/groupbutler_bot/}.
\newblock Accessed on 2024-02-01.

\bibitem{LLM_chatbot}
BYBY.DEV.
\newblock Top 8 llm-powered ai chatbots.
\newblock \url{https://byby.dev/ai-chatbots}, Nov 2023.
\newblock Accessed on 2024-02-07.

\bibitem{chen2020anonymous}
Kaiming Chen and Jiageng Chen.
\newblock Anonymous end to end encryption group messaging protocol based on asynchronous ratchet tree.
\newblock In {\em Information and Communications Security}, pages 588--605. Springer, 2020.

\bibitem{chen2022experimental}
Yunang Chen, Yue Gao, Nick Ceccio, Rahul Chatterjee, Kassem Fawaz, and Earlence Fernandes.
\newblock Experimental security analysis of the app model in business collaboration platforms.
\newblock In {\em 31st USENIX Security Symposium (USENIX Security 22)}, pages 2011--2028, 2022.

\bibitem{chia2012app}
Pern~Hui Chia, Yusuke Yamamoto, and N~Asokan.
\newblock Is this app safe? a large scale study on application permissions and risk signals.
\newblock In {\em Proceedings of the 21st international conference on World Wide Web}, pages 311--320, 2012.

\bibitem{go-mls}
Cisco.
\newblock go-mls.
\newblock \url{https://github.com/cisco/go-mls/}.
\newblock Accessed on 2024-02-08.

\bibitem{3rdparty_whatsapp_bot}
codigoencasa.
\newblock Chatbot library.
\newblock \url{https://github.com/codigoencasa/bot-whatsapp?tab=readme-ov-file}.
\newblock Accessed on 2024-02-07.

\bibitem{cohn2020formal}
Katriel Cohn-Gordon, Cas Cremers, Benjamin Dowling, Luke Garratt, and Douglas Stebila.
\newblock A formal security analysis of the signal messaging protocol.
\newblock {\em Journal of Cryptology}, 33:1914--1983, 2020.

\bibitem{cohn2016post}
Katriel Cohn-Gordon, Cas Cremers, and Luke Garratt.
\newblock On post-compromise security.
\newblock In {\em 2016 IEEE 29th Computer Security Foundations Symposium (CSF)}, 2016.

\bibitem{cohn2018ends}
Katriel Cohn-Gordon, Cas Cremers, Luke Garratt, Jon Millican, and Kevin Milner.
\newblock On ends-to-ends encryption: Asynchronous group messaging with strong security guarantees.
\newblock In {\em Proceedings of the 2018 ACM SIGSAC Conference on Computer and Communications Security}, pages 1802--1819, 2018.

\bibitem{LINEprotocol}
LY~Corporation.
\newblock Messaging api overview.
\newblock \url{https://developers.line.biz/en/docs/messaging-api/overview/}.
\newblock Accessed on 2024-01-29.

\bibitem{Platform_usage_popularity}
David Curry.
\newblock Messaging app revenue and usage statistics (2024).
\newblock \url{https://www.businessofapps.com/data/messaging-app-market/}, Jan 2024.
\newblock Accessed on 2024-02-07.

\bibitem{discord_interactions}
Discord.
\newblock Discord developer portal — documentation — application commands.
\newblock \url{https://discord.com/developers/docs/interactions/application-commands}.
\newblock Accessed on 2024-09-04.

\bibitem{discord_review_policy}
Discord.
\newblock Message content intent review policy – developers.
\newblock \url{https://support-dev.discord.com/hc/en-us/articles/5324827539479-Message-Content-Intent-Review-Policy#h_01FJ9G3JV1H0HN25C4V7X3RHD4}.
\newblock Accessed on 2024-09-03.

\bibitem{drweb}
Dr.Web.
\newblock Dr.web bot for telegram.
\newblock \url{https://free.drweb.com/drweb+telegram/}.
\newblock Accessed on 2024-02-07.

\bibitem{dubois2020speakers}
Daniel~J Dubois, Roman Kolcun, Anna~Maria Mandalari, Muhammad~Talha Paracha, David Choffnes, and Hamed Haddadi.
\newblock When speakers are all ears: Characterizing misactivations of iot smart speakers.
\newblock {\em Proceedings on Privacy Enhancing Technologies}, 2020.

\bibitem{edu2022exploring}
Jide Edu, Cliona Mulligan, Fabio Pierazzi, Jason Polakis, Guillermo Suarez-Tangil, and Jose Such.
\newblock Exploring the security and privacy risks of chatbots in messaging services.
\newblock In {\em Proceedings of the 22nd ACM internet measurement conference}, 2022.

\bibitem{emura2022membership}
Keita Emura, Kaisei Kajita, Ryo Nojima, Kazuto Ogawa, and Go~Ohtake.
\newblock Membership privacy for asynchronous group messaging.
\newblock In {\em International Conference on Information Security Applications}, pages 131--142. Springer, 2022.

\bibitem{felt2011android}
Adrienne~Porter Felt, Erika Chin, Steve Hanna, Dawn Song, and David Wagner.
\newblock Android permissions demystified.
\newblock In {\em Proceedings of the 18th ACM Conference on Computer and Communications Security}, CCS '11, page 627–638, New York, NY, USA, 2011. Association for Computing Machinery.

\bibitem{felt2012android}
Adrienne~Porter Felt, Elizabeth Ha, Serge Egelman, Ariel Haney, Erika Chin, and David Wagner.
\newblock Android permissions: User attention, comprehension, and behavior.
\newblock In {\em Proceedings of the eighth symposium on usable privacy and security}, pages 1--14, 2012.

\bibitem{gao2019autoper}
Hongcan Gao, Chenkai Guo, Yanfeng Wu, Naipeng Dong, Xiaolei Hou, Sihan Xu, and Jing Xu.
\newblock Autoper: Automatic recommender for runtime-permission in android applications.
\newblock In {\em 2019 IEEE 43rd Annual Computer Software and Applications Conference (COMPSAC)}, volume~1, pages 107--116. IEEE, 2019.

\bibitem{gieselmann2023more}
Miriam Gieselmann and Kai Sassenberg.
\newblock The more competent, the better? the effects of perceived competencies on disclosure towards conversational artificial intelligence.
\newblock {\em Social Science Computer Review}, 41(6):2342--2363, 2023.

\bibitem{gumusel2024literature}
Ece Gumusel.
\newblock A literature review of user privacy concerns in conversational chatbots: A social informatics approach: An annual review of information science and technology (arist) paper.
\newblock {\em Journal of the Association for Information Science and Technology}, 2024.

\bibitem{gunther1990identity}
Christoph~G G{\"u}nther.
\newblock An identity-based key-exchange protocol.
\newblock In {\em Advances in Cryptology --- EUROCRYPT '89}, pages 29--37. Springer Berlin Heidelberg, 1990.

\bibitem{hashimoto2022hide}
Keitaro Hashimoto, Shuichi Katsumata, and Thomas Prest.
\newblock How to hide metadata in mls-like secure group messaging: simple, modular, and post-quantum.
\newblock In {\em Proceedings of the 2022 ACM SIGSAC Conference on Computer and Communications Security}, pages 1399--1412, 2022.

\bibitem{MTProto}
Telegram~Messenger Inc.
\newblock Mtproto mobile protocol.
\newblock \url{https://core.telegram.org/mtproto}.
\newblock Accessed on 2024-01-29.

\bibitem{telegram2023api}
Telegram~Messenger Inc.
\newblock Telegram bot api.
\newblock \url{https://core.telegram.org/bots/api}.
\newblock Accessed on 2024-02-07.

\bibitem{ischen2020privacy}
Carolin Ischen, Theo Araujo, Hilde Voorveld, Guda van Noort, and Edith Smit.
\newblock Privacy concerns in chatbot interactions.
\newblock In {\em Chatbot Research and Design: Third International Workshop, CONVERSATIONS 2019, Amsterdam, The Netherlands, November 19--20, 2019, Revised Selected Papers 3}, pages 34--48. Springer, 2020.

\bibitem{keybasebasic}
Keybase.
\newblock End-to-end encryption for things that matter.
\newblock \url{https://keybase.io/}.
\newblock Accessed on 2024-06-10.

\bibitem{keybasebots}
Keybase.
\newblock Introducing bots on keybase.
\newblock \url{https://keybase.io/blog/bots}.
\newblock Accessed on 2024-06-10.

\bibitem{keybasechat}
Keybase.
\newblock Keybase book: Keybase docs.
\newblock \url{https://book.keybase.io/docs/chat/}.
\newblock Accessed on 2024-09-03.

\bibitem{kim2004tree}
Yongdae Kim, Adrian Perrig, and Gene Tsudik.
\newblock Tree-based group key agreement.
\newblock {\em ACM Transactions on Information and System Security (TISSEC)}, 7(1):60--96, 2004.

\bibitem{lau2018alexa}
Josephine Lau, Benjamin Zimmerman, and Florian Schaub.
\newblock Alexa, are you listening? privacy perceptions, concerns and privacy-seeking behaviors with smart speakers.
\newblock {\em Proceedings of the ACM on human-computer interaction}, 2(CSCW):1--31, 2018.

\bibitem{line2021encryption}
{LINE Corporation}.
\newblock Technical whitepaper - line encryption overview.
\newblock \url{https://d.line-scdn.net/stf/linecorp/en/csr/line-encryption-whitepaper-ver2.1.pdf}, Nov 2021.
\newblock Accessed on 2024-02-07.

\bibitem{LINEtransparency}
{LINE Corporation}.
\newblock Line transparency report.
\newblock \url{https://linecorp.com/en/security/encryption/2021h1}, Jan 2022.
\newblock Accessed on 2024-02-07.

\bibitem{whatsapp_webjs}
Pedro~S. Lopez.
\newblock whatsapp-web.js.
\newblock \url{https://github.com/pedroslopez/whatsapp-web.js}.
\newblock Accessed on 2024-02-08.

\bibitem{manikonda2018s}
Lydia Manikonda, Aditya Deotale, and Subbarao Kambhampati.
\newblock What's up with privacy? user preferences and privacy concerns in intelligent personal assistants.
\newblock In {\em Proceedings of the 2018 AAAI/ACM Conference on AI, Ethics, and Society}, pages 229--235, 2018.

\bibitem{perrin2016double}
Trevor Perrin and Moxie Marlinspike.
\newblock The double ratchet algorithm.
\newblock \url{https://signal.org/docs/specifications/doubleratchet/}, Nov 2016.
\newblock Accessed on 2024-02-07.

\bibitem{rosler2018more}
Paul R{\"o}sler, Christian Mainka, and J{\"o}rg Schwenk.
\newblock More is less: on the end-to-end security of group chats in signal, whatsapp, and threema.
\newblock In {\em 2018 IEEE European Symposium on Security and Privacy (EuroS\&P)}, pages 415--429. IEEE, 2018.

\bibitem{keybasesecurity}
Keegan Ryan, Thomas Pornin, and Shawn Fitzgerald.
\newblock Protocol security review.
\newblock \url{https://keybase.io/docs-assets/blog/NCC_Group_Keybase_KB2018_Public_Report_2019-02-27_v1.3.pdf}, 2019.

\bibitem{schonherr2022exploring}
Lea Sch{\"o}nherr, Maximilian Golla, Thorsten Eisenhofer, Jan Wiele, Dorothea Kolossa, and Thorsten Holz.
\newblock Exploring accidental triggers of smart speakers.
\newblock {\em Computer Speech \& Language}, 73:101328, 2022.

\bibitem{libsignal}
Signal.
\newblock libsignal.
\newblock \url{https://github.com/signalapp/libsignal/tree/main/rust/protocol/src}.
\newblock Accessed on 2024-02-08.

\bibitem{privategroup}
Signal.
\newblock Private group messaging.
\newblock \url{https://signal.org/blog/private-groups/}, 2014.
\newblock Accessed on 2024-09-05.

\bibitem{slack2023permission}
Slack.
\newblock Permission scopes.
\newblock \url{https://api.slack.com/scopes}.
\newblock Accessed on 2024-01-29.

\bibitem{slack_whitepaper}
Slack.
\newblock Security at slack.
\newblock \url{https://a.slack-edge.com/964df/marketing/downloads/security/Security_White_Paper_2020.pdf}.
\newblock Accessed on 2024-01-30.

\bibitem{staab2023beyond}
Robin Staab, Mark Vero, Mislav Balunovi{\'c}, and Martin Vechev.
\newblock Beyond memorization: Violating privacy via inference with large language models.
\newblock {\em arXiv preprint arXiv:2310.07298}, 2023.

\bibitem{subash2022disco}
Keerthana~Muthu Subash, Lakshmi~Prasanna Kumar, Sri~Lakshmi Vadlamani, Preetha Chatterjee, and Olga Baysal.
\newblock Disco: A dataset of discord chat conversations for software engineering research.
\newblock In {\em Proceedings of the 19th International Conference on Mining Software Repositories}, pages 227--231, 2022.

\bibitem{telegram2023privacy2}
{Telegram Messenger Inc.}
\newblock Telegram privacy policy.
\newblock \url{https://telegram.org/privacy}, Apr 2023.
\newblock Accessed on 2024-02-07.

\bibitem{Telegramcloudchats}
{Telegram Messenger Inc.}
\newblock Telegram support: End-to-end encryption faq.
\newblock \url{https://tsf.telegram.org/manuals/e2ee-simple}, 2023.
\newblock Accessed on 2024-02-07.

\bibitem{yandexbot}
Yandex Translate.
\newblock Telegram bot.
\newblock \url{https://yandex.com/support/translate-mobile/bot.html}.
\newblock Accessed on 2024-02-07.

\bibitem{unger2015sok}
Nik Unger, Sergej Dechand, Joseph Bonneau, Sascha Fahl, Henning Perl, Ian Goldberg, and Matthew Smith.
\newblock Sok: Secure messaging.
\newblock In {\em 2015 IEEE Symposium on Security and Privacy}, pages 232--249, 2015.

\bibitem{Viber_Whitepaper}
Rakuten Viber.
\newblock Viber encryption overview.
\newblock \url{https://www.viber.com/app/uploads/viber-encryption-overview.pdf}.
\newblock Accessed on 2024-02-04.

\bibitem{WhatsApp}
WhatsApp.
\newblock Whatsapp encryption overview.
\newblock \url{https://www.whatsapp.com/security/WhatsApp-Security-Whitepaper.pdf}, Jan 2023.
\newblock Accessed on 2024-02-07.

\bibitem{wijesekera2018contextualizing}
Primal Wijesekera, Joel Reardon, Irwin Reyes, Lynn Tsai, Jung-Wei Chen, Nathan Good, David Wagner, Konstantin Beznosov, and Serge Egelman.
\newblock Contextualizing privacy decisions for better prediction (and protection).
\newblock In {\em Proceedings of the 2018 CHI Conference on Human Factors in Computing Systems}, pages 1--13, 2018.

\end{thebibliography}
